\documentclass[12pt]{article}

\usepackage[totalheight = 23cm, totalwidth = 17cm]{geometry}
\usepackage{amssymb,amsmath,amsfonts,amsbsy,graphicx,bm}

\usepackage{color,cancel,ulem}

\newcommand{\mpl}{m_{\rm Pl}}
\newcommand{\calA}{{\cal A}}
\newcommand{\calC}{{\cal C}}
\newcommand{\calH}{{\cal H}}
\newcommand{\calO}{{\cal O}}

\begin{document}

\begin{titlepage}

\begin{center}

{\LARGE \bf 
Quantum nature of Wigner function for \\ \vspace{0.25em} inflationary tensor perturbations
}

\vskip 1.0cm

{\large
Jinn-Ouk Gong$^{a}$ 
and 
Min-Seok Seo$^{b}$ 
}

\vskip 0.5cm

{\it
$^{a}$Korea Astronomy and Space Science Institute, Daejeon 34055, Korea
\\
$^{b}$Department of Physics Education, Korea National University of Education
\\ 
Cheongju 28173, Korea
}

\vskip 1.2cm

\end{center}

\begin{abstract}

We study the Wigner function for the inflationary tensor perturbation defined in the real phase space. We compute explicitly the Wigner function including the contributions from the cubic self-interaction Hamiltonian of tensor perturbations. Then we argue that it is no longer an appropriate description for the probability distribution in the sense that quantum nature allows negativity around vanishing phase variables. This comes from the non-Gaussian wavefunction in the mixed state as a result of the non-linear interaction between super- and sub-horizon modes. We also show that this is related to the explicit infrared divergence in the Wigner function, in contrast to the trace of the density matrix.

\end{abstract}

\end{titlepage}

\newpage

\section{Introduction}

Quantum mechanics has been established as a framework for describing nature down to the microscopic scale, yet there still remain many conceptual questions to be answered. In particular, a large gap between the classical and quantum description has brought about the attempts to explain the quantum-to-classical transition dynamically. Remarkably, cosmology is expected to provide a natural testing ground for this issue (for a recent review, see \cite{Martin:2019wta}). That comes from the fact that, whereas the primordial fluctuations as we observe from the cosmic microwave background are classical,  it is likely to be originated from the quantum fluctuations \cite{Mukhanov:1981xt, Guth:1982ec, Hawking:1982cz, Starobinsky:1982ee, Bardeen:1983qw} as supported by the inflationary cosmology \cite{Guth:1980zm, Linde:1981mu, Albrecht:1982wi}.

In order to find out the observable that reveals the quantum origin of the primordial fluctuations, we need to make clear the meaning of the quantum-to-classical transition in the cosmological context. Since the typical size of the interaction between gravitons or curvature perturbation is given by $H/\mpl \ll 1$ or even more suppressed by the factor of the slow-roll parameter, the dominant loss of the quantum nature can be discussed in terms of the quadratic action only. More concretely, the quadratic action encodes the incessant interaction between the quantum fluctuations and  gravitational background as well. Then a set of  quantum fluctuations behaves coherently, forming the two-mode squeezed state about which the quantum effects coming from the non-commutativity of field operators are suppressed \cite{Guth:1985ya, Grishchuk:1989ss, Grishchuk:1990bj, Albrecht:1992kf, Polarski:1995jg}. Here the two-mode means that the quantum fluctuation with three-momentum ${\bm k}$ and that with $-{\bm k}$ appear pairwise. But still, such a squeezing is a unitary evolution which transforms the pure state into the pure state. Hence we need an additional mechanism that converts the pure state into the mixed state, in which the (diagonalized) density matrix elements correspond to the quantum analog of classical probability distribution. This is achieved by decoherence, in which the system and the environment interact with each other through higher-order non-linear interaction (for reviews, see \cite{Zurek:2003zz, Schlosshauer:2003zy}). In the inflationary universe, the horizon of size $1/H$ becomes a natural cutoff distinguishing the system from the environment: an effective field theory below the energy scale $H$ for the super-horizon modes is an open system that interacts with the sub-horizon modes as an environment \cite{Burgess:2006jn, Kiefer:2006je, Burgess:2014eoa, Burgess:2015ajz, Nelson:2016kjm, Shandera:2017qkg, Martin:2018zbe, Gong:2019yyz}. Then the evolution of the super-horizon modes is not unitary, as characterized by the Lindblad terms \cite{Lindblad:1975ef}. They describe the transition from the pure to the mixed state in decoherence through the time evolution of the density matrix \cite{Banks:1983by}. One of natural choices of the basis for the density matrix is a set of states obtained from the unitary time evolution of different particle number states. Since the unitary time evolution includes squeezing, each of basis states is the coherent superposition of multiparticle states, but still forms an orthogonal set of basis. Then we have the classical probability that the universe has evolved from different initial states, in which some of quantum fluctuations have excited \cite{Gong:2019yyz}.

With the mechanism described above, one may naively expect that cosmological perturbations as we observe are completely described in terms of classical physics. However, unlike the name stands, such a ``quantum-to-classical transition'' still preserves the quantum nature, from which we may confirm the quantum origin of cosmological perturbations. For example, whereas the density matrix is a quantum analog of the classical probability distribution, they are represented in terms of the quantum states we observe\footnote{
Since the quantum interference effects disappear through decoherence, one may attempt to explain the collapse of wavefunction after measurement in terms of it. However, the probabilistic nature still remains, so we need to employ another interpretation scheme in addition, such as the many-world interpretation. For more discussion, see, e.g. \cite{Schlosshauer:2003zy} and references therein. 
}. On the other hand, in statistical mechanics, the classical probability distribution is defined in the phase space: it is a function of canonical variables which are eigenvalues of  non-commuting quantum operators. The Wigner (distribution) function \cite{Wigner:1932eb} is devised to construct the classical probability distribution of such type from the density matrix (see \cite{Habib:1990hz, Habib:1990hx, Martin:2015qta} for early discussions in the context of cosmology). However, only limited system allows the Wigner function to be interpreted as the probability distribution since its positivity is not guaranteed \cite{Hudson:1974}. The Wigner function is positive definite only when the states comprising the density matrix is represented by the Gaussian wavefunction. In the case of the harmonic oscillator, for example, since the Gaussian wavefunction appears only in the ground state, the Wigner function is positive definite only if we have the pure ground state. Regarding the squeezed state, the squeezing of the vacuum state is represented by the Gaussian wavefunction but when the squeezed states of the initial quantum excitations are mixed, we can find the region in the phase space in which the Wigner function becomes negative. This shows that the probability distribution in the phase space is not always well-defined, reflecting the quantum nature of the primordial fluctuations even after decoherence. Therefore, the negativity of the Wigner function is a signal of quantum origin of the system. We also note that the positivity of the Wigner function does not guarantee the absence of the quantum effect\footnote{We thank referee for pointing out this.} \cite{Revzen:2004, Martin:2017zxs}. For instance, we may find out a Bell inequality violation even if the Wigner function is non-negative. Checking the positivity of the Wigner function is just a basic test for the quantum nature of the system.

In this article, we study the appearance of such negativity of the Wigner function in the context of cosmological perturbations. Our conclusion is that the higher-order non-linear interaction gives rise to the possibility for our universe to have the non-Gaussian wavefunction or equivalently, the squeezed states of the initial quantum excitation through the super- and sub-horizon interaction, hence the absolute positivity of the Wigner function is not guaranteed. Moreover, the negative Wigner function is related to its infrared divergence. This is in contrast to the unit value of the trace of the density matrix, which is the result of the cancellation between the infrared divergence in the no-graviton-excitation state and that in the soft graviton states. Such an explicit infrared divergence implies that the Wigner function is no longer a meaningful description for the probability distribution of the cosmological perturbations. For this purpose, we consider the tensor perturbations, or the graviton fluctuations, for more quantitative analysis. Whereas the cosmological perturbations also contain the curvature perturbation, reflecting a spontaneous breaking of the de Sitter (dS) isometry in quasi-dS spacetime \cite{Cheung:2007st, Prokopec:2010be, Gong:2016qpq, Gong:2017wgx}, the mechanism for the negativity of the Wigner function is essentially the same as that of the tensor perturbations: the only qualitative difference is a factor of the slow-roll parameter  which is an order parameter for the breaking of the dS isometry. In Section \ref{sec:state}, we briefly review the squeezing of tensor perturbations, which is useful in our discussions. From this, in Section \ref{eq:wigner}, we present the Wigner function explicitly to show the breaking of its positivity as a result of nonlinear interaction  which was intensively studied in \cite{Gong:2019yyz}. Implication of the Wigner function is visited in Section \ref{sec:discussions} after which we conclude. We also provide appendix sections to give some details not presented in the main text.

\section{Quantum state for tensor perturbations}
\label{sec:state}

We consider the spatial metric as
\begin{equation}
g_{ij} = a^2(\tau) ( \delta_{ij} + h_{ij} ) 
\, ,
\end{equation}
where $d\tau = dt/a$ is the conformal time and $h_{ij}$ is the pure tensor perturbations with $h^i{}_i = \partial_ih^i{}_j = 0$. Since there are two physical degrees of freedom for $h_{ij}$, we introduce the polarization tensor $e_{ij}(\lambda)$, with $\lambda$ being the polarization index, so that $h_{ij} = \sum_{\lambda=1}^2 h_\lambda e_{ij}(\lambda)$. For canonical normalization, we introduce
\begin{equation}
v_\lambda \equiv \frac{a\mpl}{\sqrt{2}} h_\lambda
\, ,
\end{equation}
then there are two copies of the identical action of a canonically normalized scalar field $v_\lambda$ for each polarization state so from now on we just consider only one polarization state and drop the subscript $\lambda$. At quadratic level, adopting the Heisenberg picture, the conjugate pair for tensor perturbations can be written as
\begin{align}
\hat{\pi}_{\bm k}(\tau)
& =
a_{\bm k}(\tau) u_k + a_{-{\bm k}}^\dag(\tau) u_k^*
\, ,
\\
\label{eq:canonical-v}
\hat{v}_{\bm k}(\tau)
& =
a_{\bm k}(\tau) v_k + a_{-{\bm k}}^\dag(\tau) v_k^*
\, ,
\end{align}
where the time-dependence is given, not to the mode functions, but to the creation and annihilation operators. Here, the initial mode functions $u_k$ and $v_k$ are found to be
\begin{equation}
u_k = - i \sqrt{\frac{k}{2}}
\quad \text{and} \quad
v_k = \frac{1}{\sqrt{2k}}
\, .
\end{equation}
What is important here is that we \textit{cannot} isolate the evolution of the pure sub-system of the mode with ${\bm k}$: when we excite the ${\bm k}$ mode, $-{\bm k}$ mode is at the same time excited as well. This becomes clear if, from the above expression, we express the creation and annihilation operators in terms of the canonical variables as
\begin{align}
\label{eq:annihilation-op}
a_{\bm k}(\tau)
& =
\frac{1}{\sqrt{2}} \Big[ \sqrt{k} \hat{v}_{\bm k}(\tau) + \frac{i}{\sqrt{k}} \hat{\pi}_{\bm k}(\tau) \Big]
\, ,
\\
\label{eq:creation-op}
a_{\bm k}^\dag(\tau)
& =
\frac{1}{\sqrt{2}} \Big[ \sqrt{k} \hat{v}_{-{\bm k}}(\tau) - \frac{i}{\sqrt{k}} \hat{\pi}_{-{\bm k}}(\tau) \Big]
\, ,
\end{align}
where it is very important to note that $a_{\bm k}^\dag(\tau)$ is given by the canonical pair of \textit{not} ${\bm k}$ but $-{\bm k}$. This is obviously because the particle creation always occurs in pair with the opposite momenta. To isolate the mode with ${\bm k}$ from that with $-{\bm k}$, we ``define'' the position $\tilde{q}_{\bm k}$ and momentum $\tilde{p}_{\bm k}$ which are purely written in terms of the operators with the corresponding momentum ${\bm k}$, {\it not} involving the opposite momentum $-{\bm k}$ as
\begin{align}
\label{eq:2modevar1tilde}
\tilde{q}_{\bm k}
& \equiv
\frac{1}{\sqrt{2k}} \Big[ a_{\bm k}(\tau) + a_{\bm k}^\dag(\tau) \Big]
\, ,
\\
\label{eq:2modevar2tilde}
\tilde{p}_{\bm k}
& \equiv
- i \sqrt{\frac{k}{2}} \Big[ a_{\bm k}(\tau) - a_{\bm k}^\dag(\tau) \Big]
\, .
\end{align}

Now we make two steps further. First, we ``discretize'' the momentum as $\int d^3k/(2\pi)^3 \to L^{-3} \sum_k$ where $L^3$ is the volume under consideration. With a given volume $L^3$, we may now isolate the volume from the canonical creation and annihilation operators $a_{\bm k}$ and $a_{\bm k}^\dag$ in such a way that, since $a_{\bm k}$ has a mass dimension of $-3/2$, we define
\begin{equation}
a_{\bm k} \equiv  L^{3/2} \hat{a}_{\bm k}
\end{equation}
and likewise for $a_{\bm k}^\dag$. That means, the new dimensionless operators $\hat{a}_{\bm k}$ and $\hat{a}^\dag_{\bm k}$ now satisfy the following commutation relation:
\begin{equation}
\Big[ \hat{a}_{\bm k}, \hat{a}^\dag_{{\bm k}'} \Big] = \delta_{{\bm k}{\bm k}'}
\, .
\end{equation}
Second, we note that $\tilde{q}_{\bm k}$ and $\tilde{p}_{\bm k}$ are of mass dimension $-2$ and $-1$ respectively. To apply directly the wisdom of the standard quantum harmonic oscillators, we rescale the dimensionful factors to introduce the dimensionless position $\hat{q}_{\bm k}$ and momentum $\hat{p}_{\bm k}$ as
\begin{align}
\label{eq:2modevar1}
\hat{q}_{\bm k} 
& \equiv
\sqrt{\frac{k}{L^3}} \tilde{q}_{\bm k}
=
\frac{1}{\sqrt{2}} \Big[ \hat{a}_{\bm k}(\tau) + \hat{a}_{\bm k}^\dag(\tau) \Big]
\, ,
\\
\label{eq:2modevar2}
\hat{p}_{\bm k}
& \equiv
\frac{1}{\sqrt{kL^3}}\tilde{p}_{\bm k}
=
- \frac{i}{\sqrt{2}} \Big[ \hat{a}_{\bm k}(\tau) - \hat{a}_{\bm k}^\dag(\tau) \Big]
\, .
\end{align}
This new pair is obviously Hermitian, i.e. $\hat{q}_{\bm k}^\dag = \hat{q}_{\bm k}$ and $\hat{p}_{\bm k}^\dag = \hat{p}_{\bm k}$, and satisfies the canonical commutation relation:
\begin{equation}
\big[ \hat{q}_{\bm k}, \hat{p}_{\bm q} \big]
=
i \big[ \hat{a}_{\bm k}, \hat{a}_{\bm q}^\dag \big]
=
i \delta_{\bm{kq}}
\, .
\end{equation}
We note that typical cosmological observables are correlators of field operators with respect to the initial vacuum, from which we can obtain information on \eqref{eq:2modevar1} and \eqref{eq:2modevar2} indirectly.  However, the Hermiticity of these operators implies that there must in principle be a clever way to ``measure'' them.

Given the canonical variables \eqref{eq:2modevar1} and \eqref{eq:2modevar2}, we can now make use of the position basis vector $|{q}_{\bm k},{q}_{-{\bm k}}\rangle$ for the two-mode state with ${\bm k}$ and $-{\bm k}$. From noting that $\widehat{H} = \sum_k \widehat{\calH}_{\bm k}$ with
\begin{equation}
\widehat{\calH}_{\bm k}
=
\frac{1}{2} \left[
k \Big( 1 + \hat{a}^\dag_{\bm k}\hat{a}_{\bm k} + \hat{a}^\dag_{-{\bm k}}\hat{a}_{-{\bm k}} \Big)
+ i \frac{a'}{a} \Big( - \hat{a}_{\bm k}\hat{a}_{-{\bm k}} + \hat{a}^\dag_{-{\bm k}}\hat{a}^\dag_{\bm k} \Big)
\right]
\, ,
\end{equation}
we can write the (free) evolution operator as a product of $\widehat{U}_{\bm k}$ which arises solely from $\widehat{\calH}_{\bm k}$:
\begin{equation}
\label{eq:freeU}
\widehat{U}_0(\tau,\tau_0)
=
\exp \left[ -i \int_{\tau_0}^\tau \widehat{H}_0(\tau') d\tau' \right]
=
\prod_i \widehat{U}_{{\bm k}_i}(\tau,\tau_0)
\, .
\end{equation}
Then the state $|\Psi\rangle$ that has evolved from the initial vacuum $|0\rangle$ at $\tau_0$,
\begin{equation}
|\Psi\rangle 
= 
\widehat{U}_0(\tau,\tau_0) |0\rangle
=
\prod_{i} \widehat{U}_{{\bm k}_i}(\tau,\tau_0) |0\rangle
\, ,
\end{equation}
using the complete basis $|{q}_{\bm k},{q}_{-{\bm k}}\rangle$, can be written 
\begin{equation}
|\Psi\rangle
=
\prod_{i} \int dq_{{\bm k}_i} dq_{-{\bm k}_i} 
\Big| {q}_{{\bm k}_i},{q}_{-{\bm k}_i} \Big\rangle \Big\langle{q}_{{\bm k}_i},{q}_{-{\bm k}_i} \Big| 
\widehat{U}_{{\bm k}_i}(\tau,\tau_0) \Big|0\Big\rangle
=
\prod_{i} \int dq_{{\bm k}_i} dq_{-{\bm k}_i} 
\Psi(q_{{\bm k}_i},q_{-{\bm k}_i}) |{q}_{{\bm k}_i},{q}_{-{\bm k}_i}\rangle
\, ,
\end{equation}
where we have defined the wavefunction $\Psi(q_{\bm k},q_{-{\bm k}})$ in the position basis $ |{q}_{\bm k},{q}_{-{\bm k}}\rangle$ as the representation of $|\Psi\rangle$. The evolution operator $\widehat{U}_{\bm k}(\tau,\tau_0)$ can be written as the product of the ``rotation'' operator $\widehat{R}_{\bm k}$ and the ``squeezing'' operator $\widehat{S}_{\bm k}$, i.e. $\widehat{U}_{\bm k} =\widehat{S}_{\bm k} \widehat{R}_{\bm k}$. Further, while the only role of $\widehat{R}_{\bm k}$ is to change the phase, $\widehat{S}_{\bm k}$ is as the name stands solely responsible for the two-mode squeezing of the initial vacuum state. More concretely, in the rotation operator
\begin{equation}
\label{eq:rotation}
\widehat{R}_{\bm k}
=
\exp \left[ -i \theta_{\bm k} \left(1+ \hat{a}_{\bm k}^\dagger \hat{a}_{\bm k} 
+ \hat{a}_{-{\bm k}}^\dagger \hat{a}_{-{\bm k}} \right)\right]
\, ,
\end{equation}
the exponents $\hat{a}_{\bm k}^\dagger \hat{a}_{\bm k}$ and $\hat{a}_{-{\bm k}}^\dagger \hat{a}_{-{\bm k}}$ are the number operators counting the number of ${\bm k}$ and $-{\bm k}$ mode excitations, respectively. Then when $\widehat{R}_{\bm k}$ is applied to $|0\rangle$, it just provides an overall phase $e^{-i\theta_{\bm k}}$. In the density matrix we will obtain in Section \ref{eq:wigner}, the phase does not play any role in the matrix element for $\widehat{U}_{\bm k}|0\rangle\langle 0|\widehat{U}_{\bm k}^\dagger=e^{-i\theta_{\bm k}}\widehat{S}_{\bm k}|0\rangle\langle 0|\widehat{S}_{\bm k}^\dagger e^{+i\theta_{\bm k}} = \widehat{S}_{\bm k}|0\rangle\langle 0|\widehat{S}_{\bm k}^\dagger$. In the same way, the phase does not contribute to the diagonal density matrix elements like $\widehat{U}_{\bm k}\hat{a}_{\bm k}^\dagger|0\rangle\langle 0|\hat{a}_{\bm k}\widehat{U}_{\bm k}^\dagger$ as well. Thus, from now on we drop the irrelevant rotation operator and consider only $\widehat{S}_{\bm k}$ in the wavefunction. The phase will be written explicitly when the off-diagonal density matrix element is discussed [see \eqref{eq:W20-result}]. The wavefunction we are interested in is given by
\begin{align}
\label{eq:2mode-wavefct}
\Psi(q_{\bm k},q_{-{\bm k}})
& =
\Big\langle {q}_{\bm k}, {q}_{-{\bm k}} \Big| \widehat{S}_{\bm k}(\tau,\tau_0) \Big|0\Big\rangle
\nonumber\\
& =
\Bigg\langle {q}_{\bm k}, {q}_{-{\bm k}} \Bigg|
\frac{1}{\cosh(r_k/2)} \sum_n \bigg[ -e^{2i\varphi_k} \tanh \bigg( \frac{r_k}{2} \bigg) \bigg]^n
\Bigg| n,{\bm k}; n,-{\bm k} \Bigg\rangle
\, ,
\end{align}
where the $n$-particle excited state, with the momenta ${\bm k}$ and $-{\bm k}$ each, is
\begin{equation}
\label{eq:n-state}
| n,{\bm k}; n,-{\bm k} \rangle
\equiv 
\frac{1}{n!} \Big( \hat{a}_{-{\bm k}}^\dag \hat{a}_{\bm k}^\dag \Big)^n |0\rangle
\, .
\end{equation}
Then it can be shown that $\Psi(q_{\bm k},q_{-{\bm k}})$ is given by
\begin{align}
\label{eq:2mode-wavefct2}
\Psi(q_{\bm k},q_{-{\bm k}})
& =
\frac{e^{A(r_k,\varphi_k) ( q_{\bm k}^2 + q_{-{\bm k}}^2 ) - B(r_k,\varphi_k)q_{\bm k}q_{-{\bm k}}}}
{\cosh(r_k/2)\sqrt{\pi}\sqrt{1-e^{4i\varphi_k}\tanh^2(r_k/2)}} 
\, ,
\\
A(r_k,\varphi_k)
& =
\frac{e^{4i\varphi_k}\tanh^2(r_k/2)+1}{2\big[ e^{4i\varphi_k}\tanh^2(r_k/2)-1 \big]} 
\, ,
\\
B(r_k,\varphi_k)
& =
\frac{-2e^{2i\varphi_k}\tanh(r_k/2)}{e^{4i\varphi_k}\tanh^2(r_k/2)-1}
\, .
\end{align}
Here, $r_k$ and $\varphi_k$ are the parameters of the Bogoliubov transformation given by \eqref{eq:dSparameter1} and \eqref{eq:dSparameter2}. The detail of the derivation of \eqref{eq:2mode-wavefct2} is given in \cite{HongYi:1989} (see also \cite{HongYi:1987zz}).

\section{Wigner function for tensor perturbations}
\label{eq:wigner}

Now we consider explicitly the reduced density matrix $\rho_\text{red}$ of the inflationary tensor perturbations on super-horizon scales under the influence of cubic interaction with sub-horizon modes. We obtain the Wigner function from $\rho_\text{red}$. If we ignore cubic or higher non-linear interactions, super- and sub-horizon modes behave independently, then the two-mode squeezed state comprised of super-horizon modes form the pure state. Since the wavefunction for the squeezed vacuum state in \eqref{eq:2mode-wavefct2} is Gaussian, the Wigner function for this state is also Gaussian and positive definite. In this case, it has been shown that the two-point correlators in terms of the quantum and the classical stochastic formalism are not distinguishable regardless of squeezing \cite{Martin:2015qta}. On the other hand, studies in this section show that when we take into account the non-linear interaction, the inevitable interaction between  super- and sub-horizon modes results in the violation of the  positivity of the Wigner function. This indicates that the Wigner function cannot be used as a conceivable probability distribution under decoherence, invalidating correlators defined upon the Wigner function as physical quantities.

Under the interaction between the super- (``system'' ${\cal S}$) and sub-horizon modes (``environment'' ${\cal E}$), tracing out the environmental sub-horizon modes introduces the non-unitary terms in the Lindblad equation:
\begin{equation}
\label{eq:Lindblad}
\frac{d}{d\tau}\rho_\text{red}(\tau)
\supset
- \frac12\sum_i \left[ L_1^\dagger L_2 \rho_\text{red}(\tau) 
+ \rho_\text{red} L_2^\dagger L_1 -2 L_1 \rho_\text{red} L_2^\dagger 
+ \left( L_1 \leftrightarrow L_2 \right) \right]
\, ,
\end{equation}
where dots indicate the irrelevant unitary evolution terms and the Lindblad operators $L_{1}$ and $L_{2}$ are defined in terms of the free, unitary evolution operator for the environment $\widehat{U}_{0, {\cal E}}$ with the subscript $0$ standing for the free evolution and the interaction Hamiltonian $H_{\rm int}$ by \cite{Shandera:2017qkg, Gong:2019yyz}
\begin{equation}
\label{eq:lindblads}
\begin{split}
L_1 & \equiv \Big\langle {\cal E}_i \Big| H_{\text{int},S}
\widehat{U}_{0,{\cal E}}(\tau;\tau_0) \Big|{\cal E}(\tau_0)\Big\rangle 
\, , 
\\
L_2 & \equiv \bigg\langle{\cal E}_i\bigg|
\int_{\tau_0}^\tau d\tau_1 H_{\text{int},I}(\tau_1-\tau)
\widehat{U}_{0,{\cal E}}(\tau;\tau_0) \bigg|{\cal E}(\tau_0)\bigg\rangle 
\, .
\end{split}
\end{equation}
Here the interaction Hamiltonian with subscripts $S$ and $I$ correspond to those written in the Schr\"odinger and interaction picture, respectively. Since the interaction Hamiltonian is dominated by cubic term for $H/\mpl \ll 1$, $L_1^\dagger L_2$ or $L_2^\dagger L_1$ in \eqref{eq:Lindblad} provide at most six creation operators for super-horizon modes depending on how many sub-horizon excitations are traced out in \eqref{eq:lindblads}. Since these excitations act on the initial vacuum, they evolve in time by the free unitary operator $\widehat{U}_{0,{\cal S}}$ dominantly. From now on, we omit the subscript ${\cal S}$ as obviously we only consider the evolution of super-horizon modes. Hence, to the lowest order in $H/\mpl$, $\rho_\text{red}$  for each super-horizon mode ${\bm k}$ is written in the schematic form
\begin{equation}
\label{eq:rhored-schematic}
\rho_\text{red}
=
\sum_{m,n\leq6} \rho_{mn} \widehat{U}_{0} a^\dag_{1} \cdots a^\dag_{m} |0\rangle
\langle0| a_{1'} \cdots a_{n} \widehat{U}^\dag_{0}
\, ,
\end{equation}
where the evolution operator, creation and annihilation operators and the vacuum state are those for the super-horizon modes. Thus we can see that $\rho_\text{red}$ contains inherently a set of basis
\begin{equation}
|a\rangle
=
\Big\{ \widehat{U}_{0}|0\rangle, \widehat{U}_{0}a^\dag_{1,}|0\rangle,
\widehat{U}_{0}a^\dag_{1}a^\dag_{2}|0\rangle, \cdots \Big\}
\, ,
\end{equation}
and using this basis can be written in the matrix form, with the subscript $ab$ denoting the element at $a$-th row and $b$-th column in the matrix, 
\begin{equation}
\label{eq:rhored}
\rho_\text{red} |_{ab}
=
\langle a | \rho_\text{red} | b \rangle
=
\begin{pmatrix}
1 - \rho_{00}  & 0 & \rho_{02} 
& \vline &
\\
0 & \rho_{11} & 0 
& \vline & 0_{3\times4}
\\
\rho_{20} & 0 & 0
& \vline &
\\
\hline
& 0_{4\times3} & 
& \vline & 0_{4\times4}
\end{pmatrix}
\, ,
\end{equation}
where $\rho_{20} = \rho_{02}^*$. 
The result is written to the leading order, ${\cal O}(H^2/\mpl^2)$. If we take the quartic or higher interactions into account, the components for the reduced density matrix will be filled more. For the detailed calculation, see \cite{Gong:2019yyz}. Note that only for the 00 component the evolution operator acts upon the vacuum state, it is the only one for which we can apply directly the wavefunction \eqref{eq:2mode-wavefct2} which, as can be seen from \eqref{eq:2mode-wavefct}, has evolved from the vacuum state. Other components such as $11$ are from the excited initial states, so require more care.

\subsubsection*{00 component}

First we consider the simplest case -- the 00 component: 
\begin{equation}
\rho_\text{red}|_{00} 
=
\big( 1 - \rho_{00} \big) \widehat{U}_{0} |0\rangle
\langle0| \widehat{U}^\dag_{0}
\, ,
\end{equation}
where the coefficient is given by \cite{Gong:2019yyz}
\begin{align}
\label{eq:rho00-coeff}
\rho_{00}
=
- \frac{36}{(2\pi)^{3}} \delta^{(3)}({\bm q}) \frac{H^2}{\mpl^{2}} 
\frac{4}{9\tau^3} 8\pi^2
\bigg( 0.577148 - \frac{32}{45}\log\varepsilon \bigg)
\, ,
\end{align}
with the infrared cutoff in the Planck unit $\varepsilon \ll 1$.

Since we are dealing with a two-mode excited state, we need to expand the simplest Wigner function \eqref{eq:wignerfct} to include the other, opposite momentum $-{\bm k}$ as \cite{Martin:2019wta}
\begin{equation}
W(q_1,q_2;p_1,p_2)
=
\int dx dy e^{-ip_1x} e^{-ip_2y}
\bigg\langle {q}_1+\frac{x}{2}, {q}_2+\frac{y}{2} \bigg|
\rho \bigg| {q}_1-\frac{x}{2}, {q}_2-\frac{y}{2} \bigg\rangle
\, .
\end{equation}
As we have noted, concentrating only on $\widehat{S}_{\bm k}$ for the evolution operator with the subscript 1 and 2 denoting respectively ${\bm k}$ and $-{\bm k}$, the contribution of $\rho_{\rm red}|_{00}$ to the Wigner function becomes
\begin{align}
\label{eq:W00-1}
W_{00}(q_{\bm k},q_{-{\bm k}};p_{\bm k},p_{-{\bm k}})
& =
\big( 1 - \rho_{00} \big) 
\int dx dy e^{-ip_{\bm k}x} e^{-ip_{-{\bm k}}y}
\Psi \bigg( q_{\bm k}+\frac{x}{2}, q_{-{\bm k}}+\frac{y}{2} \bigg)
\Psi^* \bigg( q_{\bm k}-\frac{x}{2}, q_{-{\bm k}}-\frac{y}{2} \bigg)
\, .
\end{align}
Note that as mentioned before we can write the Wigner function in terms of the wavefunction only for the 00 component. After some calculations, with the detail given in Appendix~\ref{app:wigner}, we find
\begin{align}
\label{eq:W00-result}
W_{00}(q_{\bm k},q_{-{\bm k}};p_{\bm k},p_{-{\bm k}})
& =
4 \big( 1 - \rho_{00} \big)  
\exp
\Big\{
- \big( p_{\bm k}^2 + p_{-{\bm k}}^2 + q_{\bm k}^2 + q_{-{\bm k}}^2 \big) \cosh{r_k}
\nonumber\\
&
\hspace{5em}
+ 2 \Big[ \big( p_{\bm k}p_{-{\bm k}} - q_{\bm k}q_{-{\bm k}} \big) \cos(2\varphi_k)
- 2 \big( p_{\bm k}q_{-{\bm k}} + p_{-{\bm k}}q_{\bm k} \big) \sin(2\varphi_k) \Big] \sinh{r_k}
\Big\}
\nonumber\\
&
\equiv
4 \big( 1 - \rho_{00} \big)  w_{\bm k}
\, .
\end{align}
The exponent for the exponential is a four-variable function $(p_{\bm k},p_{-{\bm k}},q_{\bm k},q_{-{\bm k}})$ and is thus hard to visualize. But nevertheless what is easy to note is that the whole exponential factor is always positive definite. Thus we conclude that the Wigner function for the 00 component \eqref{eq:W00-result} is always positive.

\subsubsection*{11 component}

Now we move to the first non-trivial part, the 11 component of $\rho_\text{red}$. Schematically, from \eqref{eq:rhored-schematic} 
\begin{equation}
\label{eq:rhored11-schematic}
\rho_\text{red}|_{11}
=
\rho_{11} \widehat{U}_{0} a^\dag_{{\bm q}_a} 
|0\rangle \langle0|
a_{{\bm q}_b} \widehat{U}^\dag_{0}
\, ,
\end{equation}
where the coefficient is given by \cite{Gong:2019yyz}  (with typo corrected)
\begin{align}
\label{eq:11coeff}
\rho_{11}
& =
- 18 \delta^{(3)}({\bm q}_{ab}) \delta_{\lambda_a\lambda_b} 
\frac{H^2}{\mpl^2} \frac{4}{9\tau^3} \frac{2\pi}{q^3} 
\times
\frac{16}{525\bar{q}^3} \bigg[
8\bar{q}^5 - 70\bar{q} + 35 \log \bigg( \frac{1+\bar{q}}{1-\bar{q}} \bigg)
\bigg]
\, ,
\end{align}
where ${\bm q}_a = -{\bm q}_b \equiv {\bm q}$ and $\bar{q} \equiv q/\calH < 1$.
Before we proceed the practical calculations, let us pause and see what this element should mean. From \eqref{eq:freeU}, we can see that only the Hamiltonian density with the momentum ${\bm q}_a$ survives, as [see \eqref{eq:squeezingop} for the operator dependence of $\widehat{U}_{0}$] 
\begin{align}
\widehat{U}_{0} a^\dag_{\bm q} |0\rangle
& \sim
\int \frac{d^3k}{(2\pi)^3} \big( a^\dag_{{\bm k}}a_{\bm k} \big) 
a^\dag_{\bm q} |0\rangle
=
\int \frac{d^3k}{(2\pi)^3} a^\dag_{\bm k} 
\Big( \big[ a_{\bm k}, a^\dag_{\bm q} \big] + a^\dag_{\bm q}a_{\bm k} \Big)
|0\rangle
=
a^\dag_{\bm q} |0\rangle
\, .
\end{align}
Thus among all ${\bm k}$-modes contained in the Hamiltonian, only the mode which has the same momentum as the external one survives. This can be more directly understood by noting that in \eqref{eq:rhored11-schematic}, the evolution operator $\widehat{U}_{0}$ is operational on the state 
\begin{equation}
\label{eq:1-pt}
a^\dag_{\bm q}|0\rangle
\equiv
|1_{\bm q}\rangle
\, ,
\end{equation}
the one-particle excited state with the momentum ${\bm q}$ that belongs to the system sector, $q < \calH$. As the evolution operator is a free one, only the mode with ${\bm q}$ is responding to $|1_{\bm q}\rangle$ and all the other modes simply disappear. From now on, without losing generality we identify the super-horizon external momentum ${\bm q}$ as ${\bm k}$.

Now let us proceed the practical calculations for the Wigner function for the 11 component. Rewriting \eqref{eq:rhored11-schematic} gives
\begin{equation}
\label{eq:rhored11-schematic2}
\rho_\text{red}|_{11}
=
\rho_{11} \widehat{S}_{\bm k} a^\dag_{\bm k} 
|0\rangle \langle0|
a_{\bm k} \widehat{S}^\dag_{\bm k}
\, ,
\end{equation}
where we have dropped the rotation operator $\widehat{R}_{\bm k}$ in the evolution operator $\widehat{U}_{\bm k} = \widehat{R}_{\bm k}\widehat{S}_{\bm k}$. Here, for the annihilation operator it should be $a_{-{\bm k}}$ due to the delta function in $\rho_{11}$ as shown in \eqref{eq:11coeff}, but as the squeezing operator $\widehat{S}_{\bm k}$ will excite both ${\bm k}$ and $-{\bm k}$ modes, we do not distinguish ${\bm k}$ and $-{\bm k}$ assigned to the operators. Since $\widehat{S}_{\bm k}$ is operating on not $|0\rangle$ but $a^\dag_{\bm k}|0\rangle$, we cannot directly apply the wavefunction \eqref{eq:2mode-wavefct} as we did for the 00 component. Instead, we have to calculate how $\widehat{S}_{\bm k}$ evolves the one-particle excited state $a^\dag_{\bm k}|0\rangle$. The detail of the calculations is given in Appendix~\ref{app:wigner}, and we find
\begin{align}
\label{eq:W11-result}
W_{11}(q_{\bm k},q_{-{\bm k}};p_{\bm k},p_{-{\bm k}})
& =
4\rho_{11} 
w_{\bm k}
\Big\{ - 1 + p_{\bm k}^2 - p_{-{\bm k}}^2 + q_{\bm k}^2 - q_{-{\bm k}}^2 
+ \big( p_{\bm k}^2 + p_{-{\bm k}}^2 + q_{\bm k}^2 + q_{-{\bm k}}^2 \big) \cosh{r_k}
\nonumber\\
&
\hspace{4.5em}
+ 2 \Big[ ( p_{\bm k}q_{-{\bm k}} + p_{-{\bm k}}q_{\bm k} ) \sin(2\varphi_k)
+ ( q_{\bm k}q_{-{\bm k}} - p_{\bm k}p_{-{\bm k}} ) \cos(2\varphi_k) \Big] \sinh{r_k} \Big\}
\, .
\end{align}
This is a complicated four-variable function of $(p_{\bm k},p_{-{\bm k}},q_{\bm k},q_{-{\bm k}})$ so not easy to visualize. But nevertheless we can collect the relevant parts. For this purpose, we note that  the exponential factor is the same as $W_{00}$ and is always positive definite. Thus, the only part we need to see if negative at all is the polynomial terms inside the curly brackets. They are, for any moderate value of the squeezing parameter $r_k$, exponentially positive. Nevertheless, they become negative for very small values of $(p_{\bm k},p_{-{\bm k}},q_{\bm k},q_{-{\bm k}})$ all around zero. We will return to this issue in Section~\ref{sec:discussions}.

\subsubsection*{20 component}

Now we move to the 20 component of $\rho_\text{red}$. Schematically, from \eqref{eq:rhored-schematic} it is
\begin{equation}
\label{eq:rhored20-schematic}
\rho_\text{red}|_{20}
=
\rho_{20} \widehat{S}_{\bm k} a^\dag_{{\bm k}_a} a^\dag_{{\bm k}_b} 
|0\rangle \langle0|
\widehat{S}^\dag_{\bm k}
\, ,
\end{equation}
where the coefficient is given by
\begin{equation}
\label{eq:20coeff}
\rho_{20} = \rho_{11} e^{2ik\tau}
\, .
\end{equation}
Again, we have dropped the rotation operator $\widehat{R}_{\bm k}$ in the evolution operator $\widehat{U}_{\bm k} = \widehat{R}_{\bm k}\widehat{S}_{\bm k}$. As before, since $\widehat{S}_{\bm k}$ is operating on not $|0\rangle$ but 
\begin{equation}
\label{eq:2-pt}
\frac{1}{2} a^\dag_{{\bm k}_a} a^\dag_{{\bm k}_b} |0\rangle 
= 
\frac{1}{2} a^\dag_{-{\bm k}} a^\dag_{\bm k} |0\rangle 
\equiv 
|2_{\bm k}\rangle
\, , 
\end{equation}
we cannot directly apply the wavefunction \eqref{eq:2mode-wavefct} as we did for the 00 component. Instead, we have to calculate how $\widehat{S}_{\bm k}$ evolves the two-particle excited state $a^\dag_{-{\bm k}} a^\dag_{\bm k} |0\rangle$. The detail is given in Appendix~\ref{app:wigner}, and we find $W_{20}$ as
\begin{align}
\label{eq:W20-result}
&
W_{20}(q_{\bm k},q_{-{\bm k}};p_{\bm k},p_{-{\bm k}})
\nonumber\\
& =
2\rho_{20} w_{\bm k}
\Bigg[
\bigg\{
2\cos(2\varphi_k)\sinh{r_k} \big( p_{\bm k}^2 + p_{-{\bm k}}^2 + q_{\bm k}^2 + q_{-{\bm k}}^2 \big)
+ 4 \cos(2\varphi_k)\sin(2\varphi_k)  (\cosh{r_k}-1)
\big( p_{-{\bm k}}q_{\bm k} + p_{\bm k} q_{-{\bm k}} \big)
\nonumber\\
& 
\hspace{5em}
+ 4 \Big[ \sin^2(2\varphi_k) + \cos^2(2\varphi_k)\cosh{r_k} \Big] 
\big( q_{\bm k}q_{-{\bm k}} - p_{\bm k} p_{-{\bm k}} \big)
\bigg\}
\nonumber\\
&
\hspace{4em}
- i
\bigg\{
2\sin(2\varphi_k)\sinh{r_k} \big( p_{\bm k}^2 + p_{-{\bm k}}^2 + q_{\bm k}^2 + q_{-{\bm k}}^2 \big)
+ 4 \Big[ \cos^2(2\varphi_k) + \sin^2(2\varphi_k)\cosh{r_k} \Big] 
\big( p_{\bm k} q_{-{\bm k}} + p_{-{\bm k}}q_{\bm k} \big)
\nonumber\\
& 
\hspace{5.5em}
+ 4\sin(2\varphi_k)\cos(2\varphi_k) \big( \cosh{r_k}-1 \big)
\big( q_{\bm k}q_{-{\bm k}} - p_{\bm k} p_{-{\bm k}} \big)
\bigg\}
\Bigg] e^{-3i\theta_k}
\, ,
\end{align}
where the factor $e^{-3i\theta_k}$ at the end comes from the rotation operator \eqref{eq:rotation} acting on $|2_{\bm k}\rangle$. Note that with the coefficient of the 02 component of $\rho_\text{red}$ being given by $\rho_{02} = \rho_{20}^*$, computing $W_{02}$ explicitly gives
\begin{align}
&
W_{02}(q_{\bm k},q_{-{\bm k}};p_{\bm k},p_{-{\bm k}})
=
W_{20}^*(q_{\bm k},q_{-{\bm k}};p_{\bm k},p_{-{\bm k}})
\, .
\end{align}

\subsubsection*{Total Wigner function}

Now we consider the total Wigner function for the reduced density matrix. To begin with, we note that the Wigner function elements \eqref{eq:W00-result}, \eqref{eq:W11-result} and \eqref{eq:W20-result} are all for a {\it single} mode with ${\bm k}$. The ``total'' Wigner function of the system of our interest should include all the modes. To see how the contributions from other momentum modes are included, let us consider simply two modes, ${\bm k}_1$ and ${\bm k}_2$:
\begin{equation}
\label{eq:rho-2sho}
\rho 
=
\rho_{0000} |0_10_2\rangle \langle0_10_2|
+ \rho_{1000} |1_10_2\rangle \langle0_10_2|
+ \rho_{1010} |1_10_2\rangle \langle1_10_2|
+ \cdots
\, ,
\end{equation}
where $|n_i\rangle$ corresponds to the $n$-particle excited state with the momentum ${\bm k}_i$, e.g. $|1_1\rangle = |1_{{\bm k}_1}\rangle$, and $\rho_{ijkl}$ is an arbitrary coefficient. Now, to compute the ``total'' Wigner function of this system, since there are two momenta\footnote{
Precisely speaking, there are two excitations each, with positive and negative momenta. Further, as there are infinitely many discrete momenta, the whole state can be written as 
\begin{equation*}
\big| (n_{{\bm k}_1}, n_{-{\bm k}_1}); (n_{{\bm k}_2}, n_{-{\bm k}_2}); \cdots \big\rangle
=
\underset{i=1}{\otimes} \big| n_{{\bm k}_i}, n_{-{\bm k}_i} \big\rangle
\, ,
\end{equation*}
so that the states given in \eqref{eq:rho-2sho} [as well as in \eqref{eq:1-pt} and \eqref{eq:2-pt}] are in fact
\begin{equation*}
|0_i\rangle
=
\big| 0_{{\bm k}_i},0_{-{\bm k}_i} \big\rangle
\, ,
\quad
|1_i\rangle
=
a^\dag_{{\bm k}_i} \big| 0_{{\bm k}_i},0_{-{\bm k}_i} \big\rangle
=
\big| 1_{{\bm k}_i},0_{-{\bm k}_i} \big\rangle
\, ,
\quad
|2_i\rangle
=
\frac{1}{2} a^\dag_{-{\bm k}_i} a^\dag_{{\bm k}_i} \big| 0_{{\bm k}_i},0_{-{\bm k}_i} \big\rangle
=
\big| 1_{{\bm k}_i},1_{-{\bm k}_i} \big\rangle
\, ,
\end{equation*}
and so on.
}, we introduce two dummy integration variables $s_1$ and $s_2$ such that
\begin{align}
\label{eq:W-2sho}
W
& =
\int ds_1 ds_2 e^{-ip_1s_1} e^{-ip_2s_2}
\bigg\langle q_1+\frac{s_1}{2}, q_2+\frac{s_2}{2} \bigg| \rho 
\bigg| q_1-\frac{s_1}{2}, q_2-\frac{s_2}{2} \bigg\rangle
\nonumber\\
& =
\int ds_1 e^{-ip_1s_1} \bigg[
\frac{\rho_{0000}}{2} \bigg\langle q_1+\frac{s_1}{2} \bigg| 0_1 \bigg\rangle 
\bigg\langle 0_1 \bigg| q_1-\frac{s_1}{2} \bigg\rangle
+ \rho_{1000} \bigg\langle q_1+\frac{s_1}{2} \bigg| 1_1 \bigg\rangle 
\bigg\langle 0_1 \bigg| q_1-\frac{s_1}{2} \bigg\rangle
\nonumber\\
&
\hspace{7em}
+ \rho_{1010} \bigg\langle q_1+\frac{s_1}{2} \bigg| 1_1 \bigg\rangle 
\bigg\langle 1_1 \bigg| q_1-\frac{s_1}{2} \bigg\rangle 
+ \cdots \bigg] 
\int ds_2 e^{-ip_2s_2} \bigg\langle q_2+\frac{s_2}{2} \bigg| 0_2 \bigg\rangle
\bigg\langle 0_2 \bigg| q_2-\frac{s_2}{2} \bigg\rangle
\nonumber\\
&
\quad
+ ({\bm k}_1 \leftrightarrow {\bm k}_2)
\, ,
\end{align}
where we have written in such a way that the Wigner function of each mode is more manifest. We should first note that when we are exciting a certain mode, the others are in their vacuum states. Thus the terms inside the square brackets, multiplied by the vacuum-vacuum component of the Wigner function for the ${\bm k}_2$ mode is the total Wigner function {\it only} for the ${\bm k}_1$ mode, except that the component for which all the modes are in the vacuum states are equally distributed, in the current case divided by two as there are two modes. This is because, as the case for which all the modes are in the vacuum is unique, this should equally contribute to the Wigner function of each mode. This suggests that for a huge number of discrete modes we should divide \eqref{eq:W00-result} by the volume of the super-horizon momentum space with the ultraviolet cutoff given by $H$, $V_\text{SH} = 4\pi H^3/3 \times 1/2$, with the factor $1/2$ being due to the pairwise appearance of ${\bm k}$ and $-{\bm k}$ in the two-mode squeezed state.

Then, \eqref{eq:W-2sho} is written as
\begin{equation}
\label{eq:W-2sho2}
W
=
\bigg[ \frac{1}{2} W_{00}(q_{{\bm k}_1},q_{-{\bm k}_1};p_{{\bm k}_1},p_{-{\bm k}_1})
+ W_{11}(q_{{\bm k}_1},q_{-{\bm k}_1};p_{{\bm k}_1},p_{-{\bm k}_1}) + \cdots \bigg] w_2
+ ({\bm k}_1 \leftrightarrow {\bm k}_2)
\, .
\end{equation}
Expanding $\rho_{20} = \rho_{11} e^{2iq\tau}$, we may just sum $W_{11}$, $W_{20}$ and $W_{02}$ to find
\begin{align}
\label{eq:Wtotal}
W_{11} + W_{20} + W_{02}
& =
W_{11} + 2 \Re \big( W_{20} \big)
\nonumber\\
& =
4 \rho_{11}  
w_{\bm k}
\bigg[
- 1 + p_{\bm k}^2 - p_{-{\bm k}}^2 + q_{\bm k}^2 - q_{-{\bm k}}^2 
+ \big( p_{\bm k}^2 + p_{-{\bm k}}^2 + q_{\bm k}^2 + q_{-{\bm k}}^2 \big) \cosh{r_k}
\nonumber\\
&
\hspace{4em}
+ 2 \Big[ ( p_{\bm k}q_{-{\bm k}} + p_{-{\bm k}}q_{\bm k} ) \sin(2\varphi_k)
+ ( q_{\bm k}q_{-{\bm k}} - p_{\bm k}p_{-{\bm k}} ) \cos(2\varphi_k) \Big] \sinh{r_k} 
\nonumber\\
&
\hspace{4em}
+ \cos(2k\tau-3\theta_{\bm k}) \Big\{
2\cos(2\varphi_k)\sinh{r_k} \big( p_{\bm k}^2 + p_{-{\bm k}}^2 + q_{\bm k}^2 + q_{-{\bm k}}^2 \big)
\nonumber \\
&
\hspace{12em}
+ 4 \cos(2\varphi_k)\sin(2\varphi_k)  (\cosh{r_k}-1)
\big( p_{-{\bm k}}q_{\bm k} + p_{\bm k} q_{-{\bm k}} \big)
\nonumber\\
& 
\hspace{12em}
+ 4 \Big[ \sin^2(2\varphi_k) + \cos^2(2\varphi_k)\cosh{r_k} \Big] 
\big( q_{\bm k}q_{-{\bm k}} - p_{\bm k} p_{-{\bm k}} \big)
\Big\}
\nonumber\\
&
\hspace{4em}
+ \sin(2k\tau-3\theta_{\bm k}) \Big\{
2\sin(2\varphi_k)\sinh{r_k} \big( p_{\bm k}^2 + p_{-{\bm k}}^2 + q_{\bm k}^2 + q_{-{\bm k}}^2 \big)
\nonumber\\
& 
\hspace{12em}
+ 4 \Big[ \cos^2(2\varphi_k) + \sin^2(2\varphi_k)\cosh{r_k} \Big] 
\big( p_{\bm k} q_{-{\bm k}} + p_{-{\bm k}}q_{\bm k} \big)
\nonumber\\
& 
\hspace{12em}
+ 4\sin(2\varphi_k)\cos(2\varphi_k) \big( \cosh{r_k}-1 \big)
\big( q_{\bm k}q_{-{\bm k}} - p_{\bm k} p_{-{\bm k}} \big)
\Big\}
\bigg]
\, .
\end{align}
Thus, \eqref{eq:W-2sho2} can be written as
\begin{equation}
W
=
4 \bigg\{ \frac{1-\rho_{00}}{2} + \rho_{11} 
\Big[ -1 + p_{{\bm k}_1}^2 - p_{-{\bm k}_1}^2 + q_{{\bm k}_1}^2 - q_{-{\bm k}_1}^2 + \cdots \Big] \bigg\} w_1 w_2
+ ({\bm k}_1 \leftrightarrow {\bm k}_2)
\, .
\end{equation}
This can be immediately extended to a large number of discrete momenta to give
\begin{equation}
W
= 4 \big( 1 - \rho_{00} \big) w_1 w_2 w_3 \cdots 
+ 4 \sum_i \rho_{11} ({\bm k}_i)
\Big[ -1 + p_{{\bm k}_i}^2 - p_{-{\bm k}_i}^2 + q_{{\bm k}_i}^2 - q_{-{\bm k}_i}^2 + \cdots \Big] 
w_1 w_2 w_3 \cdots 
+ \text{all perms}
\, .
\end{equation}

Now using the de Sitter solutions for $r_k$ and $\varphi_k$ given respectively by \eqref{eq:dSparameter1} and \eqref{eq:dSparameter2}, we find
\begin{equation}
\begin{split}
\sinh{r_k}
& =
\frac{1}{2k\tau}
\, ,
\qquad\qquad\quad
\cosh{r_k}
=
\sqrt{1+\frac{1}{4k^2\tau^2}} 
\, ,
\\
\sin(2\varphi_k)
& =
\frac{2}{\sqrt{4+\dfrac{1}{k^2\tau^2}}}
\, ,
\quad
\cos(2\varphi_k)
=
\frac{1}{k\tau\sqrt{4+\dfrac{1}{k^2\tau^2}}} 
\, ,
\end{split}
\end{equation} 
then it is trivial to find the explicit time dependence. Especially, in the late-time limit $k\tau\to0$, we find
\begin{align}
&
W_{11} + W_{20} + W_{02}
\nonumber\\
&
\underset{|k\tau|\ll1}{\longrightarrow}
4\rho_{11} \exp \bigg\{ \frac{1}{2k\tau} \Big[ ( p_{\bm k} - p_{-{\bm k}} )^2 + ( q_{\bm k} + q_{-{\bm k}} )^2 \Big]
+ 4 ( p_{\bm k} q_{-{\bm k}} + p_{-{\bm k}} q_{\bm k} ) + \calO(k\tau) \bigg\}
\nonumber\\
&
\hspace{3em}
\times
\bigg\{ - \frac{3}{2k\tau} \Big[ ( p_{\bm k} - p_{-{\bm k}} )^2 + ( q_{\bm k} + q_{-{\bm k}} )^2 \Big] 
- 1 - 6 ( p_{\bm k} q_{-{\bm k}} + p_{-{\bm k}} q_{\bm k} ) 
+ p_{\bm k}^2 - p_{-{\bm k}}^2 + q_{\bm k}^2 - q_{-{\bm k}}^2 + \calO(k\tau) \bigg\}
\, .
\end{align}
This shows that for non-zero value of $(q_{\bm k}, q_{-\bm k}, p_{\bm k}, p_{-\bm k})$, $W_{11}+W_{20}+W_{02}$ becomes large and positive at late time, $-k\tau \to 0$, as the first term in the curly bracket is dominant. On the other hand, as all the value of $(q_{\bm k}, q_{-\bm k}, p_{\bm k}, p_{-\bm k})$ approach zero, only the second term, $-1$ in the curly bracket remains, which results in the negative Wigner function.

\section{Discussions}
\label{sec:discussions}

In the previous section, we have computed the Wigner function contributions from the non-linearly evolved elements in the reduced density matrix of tensor perturbations. While the exponential factor, which the 00 component also contains, remains positive definite, the factor $-1$ in the polynomial terms in \eqref{eq:Wtotal} can make the Wigner function negative for small enough values of $(p_{\bm k},p_{-{\bm k}},q_{\bm k},q_{-{\bm k}})$. One may then argue naively as follows. These variables are written essentially in terms of the mode function and its time derivative. In any reasonable model of inflation, typically the mode function becomes frozen on super-horizon scales and accordingly its time derivative almost vanishes. Further, the mode function amplitude contains the factor $H/\mpl \ll 1$. Thus, very naively, it seems that we are naturally led to have vanishingly small $(p_{\bm k},p_{-{\bm k}},q_{\bm k},q_{-{\bm k}})$.
Interestingly, Wigner function becomes negative around this region, which signals the quantum nature of the cosmological perturbation still remains even after the horizon excape.

Before getting into detail, we make one point clear. For this purpose, let us consider the Bogoliubov transformation \eqref{eq:bogoliubov}, where the time- and momentum-dependent coefficients $\alpha_k(\tau)$ and $\beta_k(\tau)$ are subject to the constraint \eqref{eq:bogoliubov-const}. The well-known ``mode function'' $v_k(\tau)$ is given by the coefficient of the initial creation and annihilation operators: from \eqref{eq:canonical-v} and \eqref{eq:bogoliubov}, we can write
\begin{align}
\hat{v}_{\bm k}
& =
a_{\bm k}(\tau) v_k + a^\dag_{-{\bm k}}(\tau) v^*_k
=
\frac{1}{\sqrt{2k}} \underbrace{ \big[ \alpha_k(\tau) + \beta^*_k(\tau) \big] }_{
= \left( 1 - \frac{i}{k\tau} \right) e^{-ik\tau}
} a_{\bm k}(\tau_0) 
+ \frac{1}{\sqrt{2k}} \big[ \alpha_k(\tau) + \beta^*_k(\tau) \big]^* a^\dag_{-{\bm k}}(\tau_0) 
\nonumber\\
& =
a_{\bm k}(\tau_0) v_k(\tau) + a^\dag_{-{\bm k}}(\tau_0) v^*_k(\tau)
\, .
\end{align}
Thus, while both $\alpha_k(\tau)$ and $\beta_k(\tau)$ contain the plane-wave $e^{\pm ik\tau}$, the mode function $v_k(\tau)$ as we know is given by a specific combination of them. Meanwhile, from \eqref{eq:2modevar1} and \eqref{eq:2modevar2}
\begin{align}
a_{\bm k}(\tau) + a^\dag_{\bm k}(\tau)
& =
\alpha_k(\tau) a_{\bm k}(\tau_0) + \beta_k(\tau) a^\dag_{-{\bm k}}(\tau_0) + c.c.
\, ,
\\
a_{\bm k}(\tau) - a^\dag_{\bm k}(\tau)
& =
\alpha_k(\tau) a_{\bm k}(\tau_0) + \beta_k(\tau) a^\dag_{-{\bm k}}(\tau_0) - c.c.
\, ,
\end{align}
we can read that 1) the position and momentum operators $\hat{q}_{\bm k}$ and $\hat{p}_{\bm q}$ have no specific combination of $\alpha_k(\tau)$ and $\beta_k(\tau)$ that gives the mode function solution $v_k(\tau)$ and thus they cannot be written in terms of $v_k(\tau)$, and 2) $\hat{q}_{\bm k}$ and $\hat{p}_{\bm q}$ contain {\it both} initial annihilation and creation operators for {\it both} ${\bm k}$ and $-{\bm k}$ modes, and thus are superpositions of $\pm{\bm k}$ modes.

Then what about the state $|{q}_{\bm k},{q}_{-{\bm k}}\rangle$? As in the case of textbook quantum harmonic oscillator, the state upon which the annihilation and creation opeartors act is the particle number state: the state that denotes 0 particle with a specific momentum ${\bm k}$ ($a_{\bm k}|0\rangle = 0$), one-particle ($a^\dag_{\bm k}|0\rangle = |1_{\bm k}\rangle$), and so on. Thus the position state $|{q}_{\bm k}\rangle$ and the particle number state $|n_{\bm k}\rangle$ -- the collection of states that can be written in terms of the vacuum $|0\rangle$ and the creation operator $a^\dag_{\bm k}$ -- are independent. What we can do is to write one state in terms of the other as, given the competeness $\sum_n |n_{\bm k}\rangle\langle n_{\bm k}| = 1$,
\begin{equation}
|{q}_{\bm k}\rangle
=
\sum_n |n_{\bm k}\rangle\langle n_{\bm k}|{q}_{\bm k}\rangle
=
\sum_n \Psi^*_n(q_{\bm k}) |n_{\bm k}\rangle
\, ,
\end{equation}
where $\Psi_n(q_{\bm k})$ is the ``wavefunction'', such as \eqref{eq:SHO-wavefunction} for one-dimensional harmonic oscillator.

Now, we can address what is the nature of ${q}_{\bm k}$ and so on that appear in \eqref{eq:Wtotal}. Inserting \eqref{eq:n-state}
\begin{align}
\tilde{q}_{\bm k} |q_{\bm k},q_{-{\bm k}} \rangle
& =
\frac{1}{\sqrt{2k}} \Big[ \alpha_k(\tau) a_{\bm k}(\tau_0) + \beta_k(\tau) a^\dag_{-{\bm k}}(\tau_0)
+ c.c. \Big] 
\sum_{n=0}^\infty |n,{\bm k};n,-{\bm k}\rangle 
\underbrace{ \langle n,{\bm k};n,-{\bm k} | q_{\bm k},q_{-{\bm k}} \rangle }_{= \Psi^*_n(q_{{\bm k}},q_{-{\bm k}})}
\nonumber\\
& =
\text{infinite sum of } |n-1;n\rangle, \, |n;n+1\rangle, \, |n+1;n\rangle, \, |n;n-1\rangle
\, .
\end{align}
Thus we conclude we find a collection of infinitely many excitations, and unlike the naive expectation we have seen at the beginning of this section, no direct relation to the mode function $v_k(\tau)$ can be found.

Since the wavefunction of the squeezed vacuum \eqref{eq:2mode-wavefct2} is Gaussian while those of the squeezed excitations are not [see \eqref{eq:wv1st} and \eqref{eq:wv2nd}], the Wigner function is not positive definite. Even worse, the Wigner function is not free of the infrared divergence. These show that Wigner function as the classical probability distribution in the phase space is not well-defined contrary to the density matrix, which is positive definite and infrared finite. To be more quantitative on the infrared finiteness, consider the Wigner function at $q_{\bm k}=q_{-{\bm k}}=0$ and $p_{\bm k}=p_{-{\bm k}}=0$, then we have
\begin{equation}
W(0,0 ; 0,0) = \int dx dy \Big\langle \frac{x}{2}, \frac{y}{2}\Big|\rho \Big| -\frac{x}{2}, -\frac{y}{2} \Big\rangle
\, .
\end{equation}
Comparing this to the trace of the density matrix which is taken over the $(q_{\bm k}, q_{-{\bm k}})$ space [we denote them by $(x, y)$ for comparison],
\begin{equation}
{\rm Tr}[\rho] = \frac14 \int dx dy \Big\langle \frac{x}{2}, \frac{y}{2} \Big| \rho 
\Big| \frac{x}{2}, \frac{y}{2} \Big\rangle
\, ,
\end{equation}
we find that the crucial difference is the additional negative sign in the ket state. For the density matrix to be well-defined description for the mixed state probability distribution, the trace of the density matrix is infrared finite by the cancellation between the infrared divergent parts of $\rho_{00}$ and $\rho_{ii}$ with $i \geq 1$. This comes from the fact that the extremely soft excitations are not distinguishable to the vacuum state, which is originally used to argue the infrared finiteness of the quantum transition amplitude \cite{Bloch:1937pw}. Then we can normalize the trace by ${\rm Tr}[\rho]=1$ reflecting the total probability is given by $1$. Explicit calculation in \cite{Gong:2019yyz} shows that at least to the order of $H^2/\mpl^2$, the infrared finiteness of the trace,
 \begin{align}
{\rm Tr}[\rho]
& = 
\big( 1 - \rho_{00} \big) \frac14 \int dx dy 
\Big\langle \frac{x}{2}, \frac{y}{2}\Big|0\Big\rangle \Big\langle 0\Big| \frac{x}{2}, \frac{y}{2} \Big\rangle 
+ \int \frac{d^3 q}{(2\pi)^3} \rho_{11} \frac14 \int dx dy 
\Big\langle \frac{x}{2}, \frac{y}{2}\Big|1_{\bm q}\Big\rangle 
\Big\langle 1_{\bm q}\Big| \frac{x}{2}, \frac{y}{2} \Big\rangle
\nonumber\\
& \quad
+ 
\int \frac{d^3 q}{(2\pi)^3} \rho_{20} \frac14 \int dx dy 
\Big\langle \frac{x}{2}, \frac{y}{2}\Big| 2_{{\bm q}, -{\bm q}}\Big\rangle 
\Big\langle 0\Big| \frac{x}{2}, \frac{y}{2} \Big\rangle + {\rm c.c.}
\nonumber\\
& =
\big( 1 - \rho_{00} \big) + \int \frac{d^3 q}{(2\pi)^3} \rho_{11}
\, ,
\end{align}
where $\int d^3q/(2\pi)^3 \rho_{11}({\bm q}) = \rho_{00}$, is well satisfied from the cancellation between infrared divergences in $\rho_{00}$ and $\rho_{11}$. Moving to $W(0,0 ; 0,0)$, since the first (second) excitation state is odd (even) under $x\to -x$ and $y \to -y$, i.e. $\langle -x, -y| 1\rangle= -\langle x, y| 1\rangle$ and $\langle -x, -y| 2\rangle= \langle x, y| 2\rangle$, we find easily 
\begin{equation}
W(0,0 ; 0,0)
=
4 \big( 1 - \rho_{00} \big) - 4 \int \frac{d^3 q}{(2\pi)^3} \rho_{11}
= 
4 \big( 1 - 2\rho_{00} \big)
\, .
\end{equation}
This can be checked explicitly from \eqref{eq:W00-result}, \eqref{eq:W11-result} and \eqref{eq:W20-result}: putting $q_{\bm k}=q_{-{\bm k}}=0$ and $p_{\bm k}=p_{-{\bm k}}=0$, we obtain respectively $W_{00}(0,0;0,0)=4 \big( 1 - \rho_{00} \big)$, $W_{11}(0,0;0,0)=-4\rho_{11}$, and $W_{20}(0,0;0,0)=0$, which is consistent with the above. The relative minus sign between two terms shows that the infrared divergence in $\rho_{11}$ is no longer cancelled with that in $\rho_{00}$ hence $W(0,0;0,0)$ cannot be infrared finite.

We can also see that if $\rho_{00} > 1/2$, $W(0,0;0,0)$ becomes negative. From \eqref{eq:W00-result}, this is possible either for large value of $H^2/\mpl^2$ or small value of $\varepsilon$. For large $H^2/\mpl^2$, however, we should expect the next-to-leading corrections in terms of the interaction Hamiltonian of ${\cal O}(H^4/\mpl^4)$ should be significant so that our perturbative results should be modified. Meanwhile, the term containing the infrared cutoff $\varepsilon$ may call for different perturbative expansion in terms of $(H^2/\mpl^2)\log\varepsilon$ and even resummed to some analytic function as can be seen in QCD. No matter how $\varepsilon$ behaves, the  point is that the infrared cutoff $\varepsilon$ is introduced to regulate the infrared divergence, which is eventually cancelled by the soft graviton contribution in $\rho_{11}$ to make Tr$[\rho]=1$ while not in the Wigner function. This indicates that the standard interpretation of the Wigner function as the classical probability in the phase space is invalid because of not only being negative, but also exhibiting ``naked'' infrared divergence.

\section{Conclusions}
\label{sec:conclusion}

In this article, we have tested whether the Wigner function can be a conceivable physical quantity measuring the probability distribution on the real phase space of the primordial tensor perturbation. We first observe that even our universe has started from the Bunch-Davies vacuum, it would eventually become the mixed state containing the squeezed states of the initial excitations as a result of the non-linear interaction. The wavefunctions of squeezed states are not Gaussian unless the Bunch-Davies vacuum is squeezed, indicating that the positivity of the Wigner function is not guaranteed. We find that around the vanishing values of the real phase space variables $(q_{\bm k}, q_{-\bm k}, p_{\bm k}, p_{-\bm k})$, the negativity of the Wigner function is evident. This in fact has to do with the cancellation of the infrared divergence to achieve Tr$[\rho]=1$. Such a cancellation does not take place in the Wigner function, which not only leads the Wigner function to the negative value, but also makes it uncontrollably divergent. These two enable us to conclude that the Wigner function is not a good description to the probability distribution of the cosmological perturbations, indicating that the quantum nature is not completely lost even after  decoherence.

\subsection*{Acknowledgements}

We thank Sugumi Kanno, J\'er\^ome Martin and Jiro Soda for discussions while this work was under progress.
JG is supported in part by the Mid-career Research Program (2019R1A2C2085023) through the National Research Foundation of Korea Research Grants.
JG also acknowledges the Korea-Japan Basic Scientific Cooperation Program supported by the National Research Foundation of Korea and the Japan Society for the Promotion of Science (2018K2A9A2A08000127), and the Asia Pacific Center for Theoretical Physics for Focus Research Program ``The origin and evolution of the Universe'' where parts of this work were presented and discussed.

\newpage

\appendix

\renewcommand{\theequation}{\Alph{section}.\arabic{equation}}

\section{Wigner function and harmonic oscillator}
\label{app:harmonicosc}
\setcounter{equation}{0}

In this section, we recall the basic of the Wigner function. Given a pair of canonical variables $(q,p)$, the Wigner function $W(q,p)$ is defined by
\begin{equation}
\label{eq:wignerfct}
W(q,p) 
\equiv 
\int_{-\infty}^\infty ds e^{-ips}
\bigg\langle q+\frac{s}{2} \bigg| \rho \bigg| q-\frac{s}{2} \bigg\rangle
\, ,
\end{equation}
where $\rho$ is the density matrix. Equivalently, by noting that $\rho \equiv |\Psi\rangle \langle\Psi|$ with $|\Psi\rangle$ being the state of the system, and that the representation of the state $|\Psi\rangle$ in the position space $x$ is given by $\Psi(x) = \langle x|\Psi\rangle$,
\begin{align}
W(q,p)
& =
\int_{-\infty}^\infty ds e^{-ips}
\bigg\langle q+\frac{s}{2} \bigg| \Psi \bigg\rangle
\bigg\langle \Psi \bigg| q-\frac{s}{2} \bigg\rangle
\nonumber\\
& =
\int_{-\infty}^\infty ds e^{-ips}
\Psi \bigg( q+\frac{s}{2} \bigg) \Psi^* \bigg( q-\frac{s}{2} \bigg)
\, .
\end{align}
Then, clearly $W(q,p)$ is real. If integrated over either position or momentum, we find
\begin{align}
\int \frac{dq}{2\pi} W(q,p)
& =
\langle p | \rho | p \rangle
\, ,
\\
\int \frac{dp}{2\pi} W(q,p)
& =
\langle q | \rho | q \rangle
\, .
\end{align}
Given a Hermitian operator $A$, we define the so-called ``Weyl transformation'' $\calA(q,p)$ as, similar to the Wigner function,
\begin{equation}
\calA(q,p)
\equiv
\int ds e^{-ips}
\bigg\langle q+\frac{s}{2} \bigg| A \bigg| q-\frac{s}{2} \bigg\rangle
\, .
\end{equation}
Integrating the product of $\calA$ and the Wigner function $W$ over $q$ and $p$ gives
\begin{equation}
\int \frac{dqdp}{2\pi} \calA(q,p) W(q,p)
=
\int dq \langle q | A\rho |q \rangle
=
\text{Tr} (A\rho)
\, .
\end{equation}

We now consider the quantum harmonic oscillator with $m = \omega = 1$. The Hamiltonian operator is written as
\begin{equation}
\widehat{H} = \frac{\hat{p}^2}{2} + \frac{\hat{q}^2}{2} \, ,
\end{equation}
where the pair $(\hat{q},\hat{p})$ is canonical:
\begin{equation}
\big[ \hat{q},\hat{p} \big] = i
\, .
\end{equation}
We can define the annihilation and creation operators respectively by 
\begin{align}
\hat{a}
& \equiv
\frac{\hat{q}+i\hat{p}}{\sqrt{2}}
\, ,
\\
\hat{a}^\dag
& \equiv
\frac{\hat{q}-i\hat{p}}{\sqrt{2}}
\, ,
\end{align}
which satisfy the canonical commutation relation:
\begin{equation}
\big[ \hat{a}, \hat{a}^\dag \big]
=
- i \big[ \hat{q}, \hat{p} \big]
= 
1 
\, .
\end{equation}
Then the normalized wavefunction in the position space $\Psi(x) = \langle x|\Psi\rangle$ is found to be
\begin{equation}
\label{eq:SHO-wavefunction}
\Psi_n(x)
=
\frac{1}{\sqrt{2^nn!}} \frac{e^{-x^2/2}}{\pi^{1/4}} H_n(x)
\, ,
\end{equation}
where $n$ is an integer and $H_n(x)$ is the Hermite polynomial.

With the wavefunction $\Psi_n(x)$, we can compute straightly the Wigner function for each state. In the table below we show the results for $n=0$, 1 and 2 explicitly:
\begin{table}[h]
\begin{center}
\makebox[0pt]{
 \begin{tabular}{c|c|c|c}
  $n$ 
  & 
  $\Psi_n(x)=\langle x|\Psi = n\rangle$ 
  & 
  $W_n(q,p)=\displaystyle\int_{-\infty}^\infty ds e^{-ips} 
  \Psi_n\left( q+\frac{s}{2} \right) \Psi_n^*\left( q-\frac{s}{2} \right)$ 
  &
  Region for $W_n(q,p) \leq 0$
  \\
  \hline\hline
  0 & $\dfrac{e^{-x^2/2}}{\pi^{1/4}}$ & $2e^{-(q^2+p^2)}$ & none
  \\
  1 & $\dfrac{\sqrt{2}}{\pi^{1/4}} x e^{-x^2/2}$ & $\Big[ 4 \big( q^2 + p^2 \big) - 2 \Big] e^{-(q^2+p^2)}$
  &
  $q^2+p^2 \leq \dfrac{1}{2}$
  \\
  2 & $\dfrac{1}{\sqrt{2}\pi^{1/4}} \big( 2x^2 - 1 \big) e^{-x^2/2}$
  &
  $2 \Big[ 2 \big( q^2 + p^2 \big)^2 - 4 \big( q^2 + p^2 \big) + 1 \Big] e^{-(q^2+p^2)}$
  &
  $1-\dfrac{1}{\sqrt{2}} \leq q^2+p^2 \leq 1+\dfrac{1}{\sqrt{2}}$
 \end{tabular}
}
\end{center}
\end{table}

\noindent
As we consider higher excited states, the corresponding Wigner functions become more complicated. But they all exhibit negative values in certain regime, not always positive definite.

\section{Bogoliubov transformation}
\label{app:bogoliubov}
\setcounter{equation}{0}

The general solutions for the Hamiltonian equations for the creation and annihilation operators are given by the linear combination of the initial ones:
\begin{equation}
\label{eq:bogoliubov}
\begin{split}
a_{\bm k}(\tau)
& =
\alpha_k(\tau) a_{\bm k}(\tau_0) + \beta_k(\tau) a^\dag_{-{\bm k}}(\tau_0)
\\
a^\dag_{-{\bm k}}(\tau)
& =
\alpha^*_k(\tau) a^\dag_{-{\bm k}}(\tau_0) + \beta^*_k(\tau) a_{\bm k}(\tau_0)
\, ,
\end{split}
\end{equation}
which is the so-called Bogoliubov transformation. Then standard commutation relations lead to the constraint $\alpha_k(\tau)$ and $\beta_k(\tau)$ are always subject to:
\begin{equation}
\label{eq:bogoliubov-const}
|\alpha_k(\tau)|^2-|\beta_k(\tau)|^2=1
\, .
\end{equation}
Then we can parametrize them in terms of the hyperbolic functions as
\begin{equation}
\label{eq:bogoliubov-sol}
\begin{split}
\alpha_k(\tau) 
& =
e^{-i\theta_k}\cosh{r_k}
\, ,
\\
\beta_k(\tau)
& =
e^{i(\theta_k + 2\varphi_k)} \sinh{r_k}
\, .
\end{split}
\end{equation}
Assuming a perfect de Sitter background so that $a'/a = -1/\tau$ , we can find analytically the solutions as \cite{Albrecht:1992kf, Polarski:1995jg} 
\begin{align}
\label{eq:dSparameter1}
r_k
& =
\sinh^{-1} \bigg( \frac{1}{2k\tau} \bigg)
\, ,
\\
\label{eq:dSparameter2}
\varphi_k
& =
\frac{\pi}{4} - \frac{1}{2} \tan^{-1} \bigg( \frac{1}{2k\tau} \bigg)
\, ,
\\
\label{eq:dSparameter3}
\theta_k
& =
k\tau + \tan^{-1} \bigg( \frac{1}{2k\tau} \bigg)
\, .
\end{align}
Especially, by plugging \eqref{eq:dSparameter1}, \eqref{eq:dSparameter2} and \eqref{eq:dSparameter3} into \eqref{eq:bogoliubov-sol}, we can find
\begin{equation}
\begin{split}
\alpha_k 
& =
\bigg( 1 - \frac{i}{2k\tau} \bigg) e^{-ik\tau}
\, ,
\\
\beta_k
& =
\frac{i}{2k\tau} e^{ik\tau}
\, ,
\end{split}
\end{equation}
which are used to derive the well-known mode function solution $v_k(\tau)$.

\section{Explicit calculations for the Wigner functions}
\label{app:wigner}
\setcounter{equation}{0}

\subsection{00 component}

We first consider the explicit calculations of $W_{00}$. From \eqref{eq:W00-1}, using the explicit form of $\Psi$ given by \eqref{eq:2mode-wavefct2},
\begin{align}
W_{00}(q_{\bm k},q_{-{\bm k}};p_{\bm k},p_{-{\bm k}})
& =
\rho_{00} 
\int dx dy e^{-ip_{\bm k}x} e^{-ip_{-{\bm k}}y}
\frac{e^{A(r_k,\varphi_k) \big[ (q_{\bm k}+x/2)^2 + (q_{-{\bm k}}+y/2)^2 \big] 
- B(r_k,\varphi_k) ( q_{\bm k}+x/2 ) ( q_{-{\bm k}}+y/2) }}
{\cosh(r_k/2)\sqrt{\pi}\sqrt{1-e^{4i\varphi_k}\tanh^2(r_k/2)}}
\nonumber\\
&
\qquad
\times 
\Bigg[ 
\frac{e^{A(r_k,\varphi_k) \big[ (q_{\bm k}-x/2)^2 + (q_{-{\bm k}}-y/2)^2 \big] 
- B(r_k,\varphi_k) ( q_{\bm k}-x/2 ) ( q_{-{\bm k}}-y/2) }}
{\cosh(r_k/2)\sqrt{\pi}\sqrt{1-e^{4i\varphi_k}\tanh^2(r_k/2)}}
\Bigg]^*
\nonumber\\
& =
\rho_{00} |\calC|^2
\int dx dy e^{-ip_{\bm k}x} e^{-ip_{-{\bm k}}y}
e^{A \big[ (q_{\bm k}+x/2)^2 + (q_{-{\bm k}}+y/2)^2 \big] 
- B ( q_{\bm k}+x/2 ) ( q_{-{\bm k}}+y/2) }
\nonumber\\
&
\hspace{5em}
\times
e^{A^* \big[ (q_{\bm k}-x/2)^2 + (q_{-{\bm k}}-y/2)^2 \big] 
- B^* ( q_{\bm k}-x/2 ) ( q_{-{\bm k}}-y/2) }
\, ,
\end{align}
where for the second equality we have used the fact that $q_{\bm k}$ is Hermitian and have defined the coefficient $|\calC|^2$ as
\begin{align}
|\calC|^2
& \equiv 
\frac{1}{\cosh(r_k/2)\sqrt{\pi}\sqrt{1-e^{4i\varphi_k}\tanh^2(r_k/2)}}
\frac{1}{\cosh(r_k/2)\sqrt{\pi}\sqrt{1-e^{-4i\varphi_k}\tanh^2(r_k/2)}}
\nonumber\\
& =
\frac{1}{\pi\sqrt{1+\sin^2(2\varphi_k)\sinh^2r_k}}
\, ,
\end{align}
which is positive definite. Further, since the exponential factor in the integral contains the mixing terms of the integration variables $x$ and $y$ as $\int dxdy \exp ( \cdots + xy + \cdots)$, the integration of the exponential factor is not in the simplest form. One typical way to separate the variables is to introduce new variables as
\begin{equation}
\begin{split}
x & \equiv \frac{1}{2}(s+t) \, ,
\\
y & \equiv \frac{1}{2}(s-t) \, ,
\end{split}
\end{equation}
then the integration measure transforms as
\begin{equation}
dxdy
= 
\bigg| \frac{\partial(x,y)}{\partial(s,t)} \bigg| dsdt
=
\frac{1}{2}dsdt
\, .
\end{equation}
Finally, we note that the exponent of \eqref{eq:2mode-wavefct2} can be written as
\begin{align}
- \frac{\alpha}{2} \big( q_{\bm k}+q_{-{\bm k}} \big)^2
- \frac{\beta}{2} \big( q_{\bm k}-q_{-{\bm k}} \big)^2
=
- \frac{1}{2}(\alpha+\beta) \big( q_{\bm k}^2 + q_{-{\bm k}}^2 \big)
- (\alpha-\beta) q_{\bm k}q_{-{\bm k}}
\, ,
\end{align}
thus we can identify $A = -(\alpha+\beta)/2$ and $B = \alpha-\beta$ so that
\begin{align}
\alpha 
& =
\frac{1+e^{2i\varphi_k}\tanh(r_k/2)}{2 \big[ 1-e^{2i\varphi_k}\tanh(r_k/2) \big]}
=
\frac{1 + i\sin(2\varphi_k)\sinh{r_k}}{2 \big[ \cosh{r_k} - \cos(2\varphi_k)\sinh{r_k} \big]}
\, ,
\\
\beta
& =
\frac{1-e^{2i\varphi_k}\tanh(r_k/2)}{2 \big[ 1+e^{2i\varphi_k}\tanh(r_k/2) \big]}
=
\frac{1 - i\sin(2\varphi_k)\sinh{r_k}}{2 \big[ \cosh{r_k} + \cos(2\varphi_k)\sinh{r_k} \big]}
\, .
\end{align}
Collecting all these, we can write $W_{00}$ analytically to find 
\begin{align}
%\label{eq:W00}
W_{00}
& =
\frac{\rho_{00}|\calC|^2}{2}
\int dsdt 
\exp \bigg\{
- \frac{1}{2} \big[ (\alpha+\alpha^*) (q_{\bm k}+q_{-{\bm k}})^2
+ (\beta+\beta^*) (q_{\bm k}-q_{-{\bm k}})^2 \big] 
\nonumber\\
&
\hspace{6em}
+ \frac{1}{2} \big[ - i (p_{\bm k}+p_{-{\bm k}}) 
- (\alpha-\alpha^*)(q_{\bm k}+q_{-{\bm k}}) \big] 
+ \frac{1}{2} \big[ - i (p_{\bm k}-p_{-{\bm k}}) 
- (\beta-\beta^*)(q_{\bm k}-q_{-{\bm k}}) \big] 
\nonumber\\
&
\hspace{6em}
- \frac{1}{8} (\alpha+\alpha^*) - \frac{1}{8} (\beta+\beta^*) 
\bigg\}
\, ,
\end{align}
which can be integrated analytically to give \eqref{eq:W00-result}.

\subsection{11 component}

We begin with the original expression for $\widehat{S}_{\bm k}$ as
\begin{align}
\label{eq:squeezingop}
\widehat{S}_{\bm k}
& =
\underbrace{ \exp \bigg[ - e^{2i\varphi_k} \tanh \bigg( \frac{r_k}{2} \bigg) \hat{a}^\dag_{-{\bm k}} 
\hat{a}^\dag_{\bm k} \bigg] }_{\equiv A}
\underbrace{ \bigg[ \frac{1}{\cosh(r_k/2)} \bigg]^{\hat{a}^\dag_{\bm k}\hat{a}_{\bm k} 
+ \hat{a}^\dag_{-{\bm k}}\hat{a}_{-{\bm k}} + 1} }_{\equiv B}
\underbrace{ \exp \bigg[ e^{-2i\varphi_k} \tanh \bigg( \frac{r_k}{2} \bigg) 
\hat{a}_{-{\bm k}} \hat{a}_{\bm k} \bigg] }_{\equiv C}
\, .
\end{align}
Operating upon $\hat{a}^\dag_{\bm k}|0\rangle $ first gives
\begin{equation}
\widehat{S}_{\bm k} \hat{a}^\dag_{\bm k}|0\rangle 
=
\Big[ \widehat{S}_{\bm k}, \hat{a}^\dag_{\bm k} \Big] |0\rangle
+ \hat{a}^\dag_{\bm k} \widehat{S}_{\bm k} |0\rangle 
\, ,
\end{equation}
and the commutator reads
\begin{equation}
\Big[ \widehat{S}_{\bm k}, \hat{a}^\dag_{\bm k} \Big]
=
\big[ ABC, \hat{a}^\dag_{\bm k} \big]
=
\big[ A, \hat{a}^\dag_{\bm k} \big] BC
+ A \big[ B, \hat{a}^\dag_{\bm k} \big] C
+ AB \big[ C, \hat{a}^\dag_{\bm k} \big]
\, .
\end{equation}
Three simplifications are ahead. First, as we can see from \eqref{eq:squeezingop}, $A$ only contains $\hat{a}^\dag_{-{\bm k}}\hat{a}^\dag_{\bm k}$, and thus the commutator with $\hat{a}^\dag_{\bm k}$ simply vanishes:
\begin{equation}
\big[ A, \hat{a}^\dag_{\bm k} \big] = 0 \, .
\end{equation}
Next, if we expand the exponential in $C$, 
\begin{equation}
\label{eq:Cexpansion}
C 
=
1
+ e^{-2i\varphi_k} \tanh \bigg( \frac{r_k}{2} \bigg) \hat{a}_{-{\bm k}} \hat{a}_{\bm k}
+ \frac{1}{2!} \bigg[ e^{-2i\varphi_k} \tanh \bigg( \frac{r_k}{2} \bigg) \bigg]^2 
\big( \hat{a}_{-{\bm k}} \hat{a}_{\bm k} \big)^2
+ \cdots
\, ,
\end{equation}
which only contains $\hat{a}_{-{\bm k}}\hat{a}_{\bm k}$. Thus, in $\big[ C, \hat{a}^\dag_{\bm k} \big]$, while one annihilation operator brings with $\hat{a}^\dag_{\bm k}$ a delta function via the canonical commutation relation, still more than one annilation operators remain which eliminate the vacuum state $|0\rangle$ multiplied to the right. Thus, we have
\begin{equation}
\label{eq:Cidentity}
\big[ C, \hat{a}^\dag_{\bm k} \big] |0\rangle = 0 \, .
\end{equation}
Finally, if $C$ is acting directly on the vacuum state, only the first term in \eqref{eq:Cexpansion} survives and the original vacuum state remains identical:
\begin{equation}
C|0\rangle = |0\rangle \, .
\end{equation}
Thus, we only have
\begin{equation}
\Big[ \widehat{S}_{\bm k}, \hat{a}^\dag_{\bm k} \Big] |0\rangle
=
A \big[ B, \hat{a}^\dag_{\bm k} \big] |0\rangle
\, .
\end{equation}

To proceed further, we consider $B$ more closely, we first single out $\hat{a}^\dag_{\bm k}\hat{a}_{\bm k}$ and expand in terms of logarithm to write
\begin{align}
\label{eq:[B,a]1}
\big[ B, \hat{a}^\dag_{\bm k} \big]
& =
\bigg[ \frac{1}{\cosh(r_k/2)} \bigg]^{\hat{a}^\dag_{-{\bm k}}\hat{a}_{-{\bm k}}+1}
\sum_{n=0}^\infty \frac{1}{n!} \bigg\{ \log \bigg[ \frac{1}{\cosh(r_k/2)} \bigg] \bigg\}^n
\Big[ \big( \hat{a}^\dag_{\bm k}\hat{a}_{\bm k} \big)^n, \hat{a}^\dag_{\bm k} \Big] 
\, .
\end{align}
Multiplying the vacuum $|0\rangle$ to the right and separately considering $n=0$,
\begin{align}
\Big[ \big( \hat{a}^\dag_{\bm k}\hat{a}_{\bm k} \big)^n, \hat{a}^\dag_{\bm k} \Big] \Big|_{n=0} |0\rangle
& =
\big[ 1, \hat{a}^\dag_{\bm k} \big] |0\rangle
= 0
\, ,
\\
\label{eq:[B,a]2}
\Big[ \big( \hat{a}^\dag_{\bm k}\hat{a}_{\bm k} \big)^n, \hat{a}^\dag_{\bm k} \Big] \Big|_{n\geq1} |0\rangle
& =
\Big[ \big( \hat{a}^\dag_{\bm k}\hat{a}_{\bm k} \big)^n \hat{a}^\dag_{\bm k} 
- \hat{a}^\dag_{\bm k} \big( \hat{a}^\dag_{\bm k}\hat{a}_{\bm k} \big)^n \Big] |0\rangle
=
\hat{a}^\dag_{\bm k} |0\rangle
\, ,
\end{align}
so that
\begin{align}
\label{eq:[B,a]3}
\big[ B, \hat{a}^\dag_{\bm k} \big] |0\rangle
=
\bigg[ \frac{1}{\cosh(r_k/2)} \bigg]^{\hat{a}^\dag_{-{\bm k}}\hat{a}_{-{\bm k}}+1}
\bigg[ \frac{1}{\cosh(r_k/2)} - 1 \bigg] \hat{a}^\dag_{\bm k} |0\rangle
\, .
\end{align}
Since $\hat{a}^\dag_{\bm k}$ and $\hat{a}^\dag_{-{\bm k}}$ commute, $\hat{a}^\dag_{\bm k}$ may go in front of the $\hat{a}^\dag_{-{\bm k}}\hat{a}_{-{\bm k}}$ term. Expanding $\hat{a}^\dag_{-{\bm k}}\hat{a}_{-{\bm k}}$ similarly gives
\begin{align}
\label{eq:[B,a]4}
\big[ B, \hat{a}^\dag_{\bm k} \big] |0\rangle
& =
\bigg[ \frac{1}{\cosh(r_k/2)} - 1 \bigg] \frac{1}{\cosh(r_k/2)} \hat{a}^\dag_{\bm k}
\bigg[ \frac{1}{\cosh(r_k/2)} \bigg]^{\hat{a}^\dag_{-{\bm k}}\hat{a}_{-{\bm k}}} |0\rangle
= 
\bigg[ \frac{1}{\cosh(r_k/2)} - 1 \bigg] \frac{1}{\cosh(r_k/2)} \hat{a}^\dag_{\bm k} |0\rangle
\, .
\end{align}
This gives a rather simple result:
\begin{align}
\Big[ \widehat{S}_{\bm k}, \hat{a}^\dag_{\bm k} \Big] |0\rangle
& =
\bigg[ \frac{1}{\cosh(r_k/2)} - 1 \bigg] \hat{a}^\dag_{\bm k}
\frac{1}{\cosh(r_k/2)} 
\sum_{n=0}^\infty \left[ - e^{2i\varphi_k} \tanh \left( \frac{r_k}{2} \right) \right]^n
\frac{1}{n!} \left( \hat{a}^\dag_{-{\bm k}} \hat{a}^\dag_{\bm k} \right)^n |0\rangle 
\nonumber\\
& 
=
\bigg[ \frac{1}{\cosh(r_k/2)} - 1 \bigg] \hat{a}^\dag_{\bm k} \widehat{S}_{\bm k}|0\rangle
\, .
\end{align}
Thus, finally,
\begin{equation}
\widehat{S}_{\bm k} \hat{a}^\dag_{\bm k}|0\rangle 
=
\Big[ \widehat{S}_{\bm k}, \hat{a}^\dag_{\bm k} \Big] |0\rangle
+ \hat{a}^\dag_{\bm k} \widehat{S}_{\bm k} |0\rangle 
=
\frac{1}{\cosh(r_k/2)} \hat{a}^\dag_{\bm k} \widehat{S}_{\bm k}|0\rangle
\, .
\end{equation}

Since now we have exchanged the position of $\hat{a}^\dag_{\bm k}$ and $\widehat{S}_{\bm k}$ such that the squeezing operator $\widehat{S}_{\bm k}$ is directly acting on the vacuum state from the left, we are in the half way of making use of the explicit form of the wavefunction $\Psi(q_{\bm k},q_{-{\bm k}}) = \Big\langle {q}_{\bm k},{q}_{-{\bm k}} \Big| \widehat{S}_{\bm k}(\tau,\tau_0) \Big|0\Big\rangle_0$ given by \eqref{eq:2mode-wavefct}. But still we have $a^\dag_{\bm k}$ lurking around, which we should work out. A hint on how to deal with $\hat{a}^\dag_{\bm k}$ comes from the explicit functional form of the wavefunction $\Psi(q_{\bm k},q_{-{\bm k}})$. It is, as explicitly shown in \eqref{eq:2mode-wavefct}, a function of $q_{\bm k}$ and $q_{-{\bm k}}$. This could well be anticipated as it is the representation of the squeezed state evolved from the vacuum $\widehat{S}_{\bm k}|0\rangle$ in the position basis $\langle {q}_{\bm k}, {q}_{-{\bm k}}|$. Thus position $q_{\bm k}$ should be our primary variable, while the conjugate momentum would be regarded as the operator in the position space, $\hat{p} = -i\partial/\partial\hat{q}$. Since \eqref{eq:2modevar1} and \eqref{eq:2modevar2} are precisely of the same form as the simplest harmonic oscillator, $\hat{a}^\dag_{\bm k}$ is written as, with $\hat{p}_{\bm k}$ being a position space operator $-i\partial/\partial\hat{q}_{\bm k}$,
\begin{equation}
\label{eq:a-op}
\hat{a}^\dag_{\bm k} 
= 
\frac{1}{\sqrt{2}} \big( \hat{q}_{\bm k} - i\hat{p}_{\bm k} \big)
=
\frac{1}{\sqrt{2}} \bigg( \hat{q}_{\bm k} - \frac{\partial}{\partial\hat{q}_{\bm k}} \bigg)
\, .
\end{equation}
We can thus write
\begin{align}
\label{eq:wv1st}
\Big\langle {q}_{\bm k},{q}_{-{\bm k}} \Big|
\widehat{S}_{\bm k} \hat{a}^\dag_{\bm k} \Big|0\Big\rangle
& =
\frac{1}{\cosh(r_k/2)}
\Big\langle {q}_{\bm k},{q}_{-{\bm k}} \Big|
\hat{a}^\dag_{\bm k} \widehat{S}_{\bm k}\Big|0\Big\rangle
\nonumber\\
& =
\frac{1}{\cosh(r_k/2)}
\frac{1}{\sqrt{2}} \bigg( {q}_{\bm k} - \frac{\partial}{\partial{q}_{\bm k}} \bigg)
\underbrace{ \Big\langle {q}_{\bm k}, {q}_{-{\bm k}} \Big|
\widehat{S}_{\bm k}\Big|0\Big\rangle }_{= \Psi(q_{\bm k},q_{-{\bm k}})}
\nonumber\\
&=
\frac{1}{\sqrt{2}\cosh(r_k/2)}
\Big[ (1- 2A)q_{\bm k} + Bq_{-{\bm k}} \Big]
\frac{e^{A(r_k,\varphi_k) ( q_{\bm k}^2 + q_{-{\bm k}}^2 ) - B(r_k,\varphi_k)q_{\bm k}q_{-{\bm k}}}}
{\cosh(r_k/2)\sqrt{\pi}\sqrt{1-e^{4i\varphi_k}\tanh^2(r_k/2)}}  
\, .
\end{align}

Now we have all the ingredients to compute the Wigner function $W_{11}$ analytically. We find explicitly
\begin{align}
\label{eq:W11}
&
W_{11}(q_{\bm k},q_{-{\bm k}};p_{\bm k},p_{-{\bm k}})
\nonumber\\
& =
\rho_{11} \frac{1}{2\cosh^2(r_q/2)} 
\int dx dy e^{-ip_{\bm k}x} e^{-ip_{-{\bm k}}y}
\bigg[ (1-2A) \bigg( q_{\bm k}+\frac{x}{2} \bigg) + B \bigg( q_{\bm k}+\frac{x}{2} \bigg) \bigg]
\Psi \bigg( {q}_{\bm k}+\frac{x}{2}, {q}_{-{\bm k}}+\frac{y}{2} \bigg)
\nonumber\\
&
\hspace{9em}
\times
\bigg[ (1-2A^*) \bigg( q_{\bm k}-\frac{x}{2} \bigg) + B^* \bigg( q_{\bm k}-\frac{x}{2} \bigg) \bigg]
\Psi^* \bigg( {q}_{\bm k}-\frac{x}{2}, {q}_{-{\bm k}}-\frac{y}{2} \bigg)
\, .
\end{align}
This is analytically integrable and then we find the result given in the main text,~\eqref{eq:W11-result}.

\subsection{20 component}

Splitting $\widehat{S}_{\bm k} = ABC$ as in \eqref{eq:squeezingop}, we find
\begin{align}
\label{eq:Saa}
\widehat{S}_{\bm k} \hat{a}^\dag_{-{\bm k}} \hat{a}^\dag_{\bm k} |0\rangle
& =
\Big[ \widehat{S}_{\bm k} , \hat{a}^\dag_{-{\bm k}} \hat{a}^\dag_{\bm k} \Big] |0\rangle
+ \hat{a}^\dag_{-{\bm k}} \hat{a}^\dag_{\bm k} \widehat{S}_{\bm k} |0\rangle
\nonumber\\
& 
=
A \big[ B, \hat{a}^\dag_{-{\bm k}} \hat{a}^\dag_{\bm k} \big] |0\rangle
+ AB \big[ C, \hat{a}^\dag_{-{\bm k}} \hat{a}^\dag_{\bm k} \big] |0\rangle
+ \hat{a}^\dag_{-{\bm k}} \hat{a}^\dag_{\bm k} \widehat{S}_{\bm k} |0\rangle
\, .
\end{align}
Thus we need to consider two new commutators:
\begin{enumerate}

\item $\big[ B, \hat{a}^\dag_{-{\bm k}} \hat{a}^\dag_{\bm k} \big] |0\rangle$

The commutator can be expanded to give
\begin{align}
\big[ B, \hat{a}^\dag_{-{\bm k}} \hat{a}^\dag_{\bm k} \big]
& =
\big[ B, \hat{a}^\dag_{-{\bm k}} \big] \hat{a}^\dag_{\bm k} 
+ \hat{a}^\dag_{-{\bm k}} \big[ B, \hat{a}^\dag_{\bm k} \big]
\, .
\end{align}
For the first term, we follow the same steps as \eqref{eq:[B,a]1} - \eqref{eq:[B,a]4}, with ${\bm k}$ being replaced by $-{\bm k}$ and accordingly related expansions in the operator $B$:
\begin{equation}
\big[ B, \hat{a}^\dag_{-{\bm k}} \big] \hat{a}^\dag_{\bm k} |0\rangle
=
\frac{1}{\cosh^2(r_k/2)}\bigg[ \frac{1}{\cosh(r_k/2)} - 1 \bigg] 
\hat{a}^\dag_{-{\bm k}} \hat{a}^\dag_{\bm k} |0\rangle
\, .
\end{equation}
Along with \eqref{eq:[B,a]4}, we have
\begin{align}
\big[ B, \hat{a}^\dag_{-{\bm k}} \hat{a}^\dag_{\bm k} \big] |0\rangle
& =
\frac{1}{\cosh^2(r_k/2)}\bigg[ \frac{1}{\cosh(r_k/2)} - 1 \bigg] 
\hat{a}^\dag_{-{\bm k}} \hat{a}^\dag_{\bm k} |0\rangle
+
\bigg[ \frac{1}{\cosh(r_k/2)} - 1 \bigg] \frac{1}{\cosh(r_k/2)} 
\hat{a}^\dag_{-{\bm k}} \hat{a}^\dag_{\bm k} |0\rangle
\nonumber\\
& =
-\frac{\tanh^2(r_k/2)}{\cosh(r_k/2)} \hat{a}^\dag_{-{\bm k}} \hat{a}^\dag_{\bm k} |0\rangle
\, .
\end{align}

\item $\big[ C, \hat{a}^\dag_{-{\bm k}} \hat{a}^\dag_{\bm k} \big] |0\rangle$

Expanding the commutator gives
\begin{equation}
\big[ C, \hat{a}^\dag_{-{\bm k}} \hat{a}^\dag_{\bm k} \big]
=
\big[ C, \hat{a}^\dag_{-{\bm k}} \big] \hat{a}^\dag_{\bm k} 
+ \hat{a}^\dag_{-{\bm k}} \big[ C, \hat{a}^\dag_{\bm k} \big]
\, .
\end{equation}
Since $\big[ C, \hat{a}^\dag_{\bm k} \big] |0\rangle = 0$, the second term vanishes. To compute the first term, from \eqref{eq:Cexpansion}
\begin{align}
\big[ C, \hat{a}^\dag_{-{\bm k}} \big] 
& =
\sum_{n=0}^\infty \frac{1}{n!} \bigg[ e^{-2i\varphi_k} \tanh \bigg( \frac{r_k}{2} \bigg) \hat{a}_{\bm k} \bigg]^n
\Big[ \big( \hat{a}_{-{\bm k}} \big)^n, \hat{a}^\dag_{-{\bm k}} \Big] 
\nonumber\\
& =
e^{-2i\varphi_k} \tanh \bigg( \frac{r_k}{2} \bigg) \hat{a}_{\bm k} C
\, ,
\end{align}
where we have used the identity $\Big[ \big( \hat{a}_{-{\bm k}} \big)^n, \hat{a}^\dag_{-{\bm k}} \Big] = n\hat{a}_{-{\bm k}}^{n-1}$. Then we find
\begin{align}
\big[ C, \hat{a}^\dag_{-{\bm k}} \hat{a}^\dag_{\bm k} \big] |0\rangle
& =
e^{-2i\varphi_k} \tanh \bigg( \frac{r_k}{2} \bigg) \hat{a}_{\bm k} 
\Big\{ \big[ C, \hat{a}^\dag_{\bm k} \big] + \hat{a}^\dag_{\bm k}C \Big\} |0\rangle
\nonumber\\
& =
e^{-2i\varphi_k} \tanh \bigg( \frac{r_k}{2} \bigg) |0\rangle
\, .
\end{align}

\end{enumerate}

Thus, \eqref{eq:Saa} can be written as
\begin{align}
\widehat{S}_{\bm k} \hat{a}^\dag_{-{\bm k}} \hat{a}^\dag_{\bm k} |0\rangle
& =
A \times -\frac{\tanh^2(r_k/2)}{\cosh(r_k/2)} \hat{a}^\dag_{-{\bm k}} \hat{a}^\dag_{\bm k} |0\rangle
+ AB e^{-2i\varphi_k} \tanh \bigg( \frac{r_k}{2} \bigg) |0\rangle
+ \hat{a}^\dag_{-{\bm k}} \hat{a}^\dag_{\bm k} \widehat{S}_{\bm k} |0\rangle
\nonumber\\
& =
\frac{1}{\cosh^2(r_k/2)} \hat{a}^\dag_{-{\bm k}} \hat{a}^\dag_{\bm k} \widehat{S}_{\bm k} |0\rangle
+ e^{-2i\varphi_k} \tanh \bigg( \frac{r_k}{2} \bigg) \widehat{S}_{\bm k} |0\rangle
\, ,
\end{align}
where we have expanded the operators in $A$ and $B$ and then resummed. Thus, multiplying the position basis $\langle {q}_{\bm k}, {q}_{-{\bm k}}|$ to the left and using \eqref{eq:a-op},
\begin{align}
\label{eq:wv2nd}
\Big\langle {q}_{\bm k}, {q}_{-{\bm k}} \Big|
\widehat{S}_{\bm k} \hat{a}^\dag_{-{\bm k}} \hat{a}^\dag_{\bm k} \Big|0\Big\rangle
& =
\frac{1}{\cosh^2(r_k/2)} \Big\langle {q}_{\bm k}, {q}_{-{\bm k}} \Big|
\hat{a}^\dag_{-{\bm k}} \hat{a}^\dag_{\bm k} \widehat{S}_{\bm k} \Big|0\Big\rangle
+ e^{-2i\varphi_k} \tanh \bigg( \frac{r_k}{2} \bigg) 
\Big\langle {q}_{\bm k}, {q}_{-{\bm k}} \Big| \widehat{S}_{\bm k} \Big|0\Big\rangle
\nonumber\\
& =
\bigg[
\frac{1}{2\cosh^2(r_k/2)} \bigg( \hat{q}_{\bm k} - \frac{\partial}{\partial\hat{q}_{\bm k}} \bigg)
\bigg( \hat{q}_{-{\bm k}} - \frac{\partial}{\partial\hat{q}_{-{\bm k}}} \bigg) 
+ e^{-2i\varphi_k} \tanh \bigg( \frac{r_k}{2} \bigg) 
\bigg] \Psi(q_{\bm k},q_{-{\bm k}})
\end{align}
Here,
\begin{align}
&
\bigg( \hat{q}_{\bm k} - \frac{\partial}{\partial\hat{q}_{\bm k}} \bigg)
\bigg( \hat{q}_{-{\bm k}} - \frac{\partial}{\partial\hat{q}_{-{\bm k}}} \bigg) \Psi(q_{\bm k},q_{-{\bm k}})
=
\Big\{ (1-2A)B \big( q_{\bm k}^2 + q_{-{\bm k}}^2 \big) 
+ \big[ (1-2A)^2+B^2 \big] q_{\bm k} q_{-{\bm k}} - B \Big\} \Psi
\, .
\end{align}
Therefore, we finally find
\begin{align}
%&
\Big\langle {q}_{\bm k}, {q}_{-{\bm k}} \Big|
\widehat{S}_{\bm k} \hat{a}^\dag_{-{\bm k}} \hat{a}^\dag_{\bm k} \Big|0\Big\rangle
& =
\frac{1}{2\cosh^2(r_k/2)} \Big\{ (1-2A)B \big( q_{\bm k}^2 + q_{-{\bm k}}^2 \big) 
+ \big[ (1-2A)^2+B^2 \big] q_{\bm k}q_{-{\bm k}} 
\nonumber\\
&
\hspace{7em}
- B + e^{-2i\varphi_k} \sinh{r_k} \Big\}
\Psi(q_{\bm k},q_{-{\bm k}}) 
\label{Eq:wv2nd}
\, .
\end{align}

Now we have all the ingredients to write the Wigner function $W_{20}$. We find explicitly
\begin{align}
&
W_{20}(q_{\bm k},q_{-{\bm k}};p_{\bm k},p_{-{\bm k}})
\nonumber\\
& =
\frac{\rho_{20}}{2\cosh^2(r_k/2)}  
\int dx dy e^{-ip_{\bm k}x} e^{-ip_{-{\bm k}}y}
\Bigg\{ (1-2A)B \left[ \bigg( q_{\bm k}+\frac{x}{2} \bigg)^2 
+ \bigg( q_{-{\bm k}}+\frac{y}{2} \bigg)^2 \right]
\nonumber\\
& 
\hspace{16em}
+ \big[ (1-2A)^2+B^2 \big] \bigg( q_{\bm k}+\frac{x}{2} \bigg) \bigg( q_{-{\bm k}}+\frac{y}{2} \bigg) - B 
+ e^{-2i\varphi_k}\sinh{r_k}
\Bigg\}
\nonumber\\
&
\hspace{15em}
\times
\Psi \bigg( {q}_{\bm k}+\frac{x}{2}, {q}_{-{\bm k}}+\frac{y}{2} \bigg)
\Psi^* \bigg( {q}_{\bm k}-\frac{x}{2}, {q}_{-{\bm k}}-\frac{y}{2} \bigg)
\, .
\end{align}
By changing the variables as $x = (s+t)/2$ and $y=(s-t)/2$ with $A = -(\alpha+\beta)/2$ and $B = \alpha-\beta$, we can perform the integrations to find \eqref{eq:W20-result}.


\begin{thebibliography}{99}





%\cite{Martin:2019wta}
\bibitem{Martin:2019wta} 
  J.~Martin,
  %``Cosmic Inflation, Quantum Information and the Pioneering Role of John S Bell in Cosmology,''
  Universe {\bf 5}, no. 4, 92 (2019)
%  doi:10.3390/universe5040092
  [arXiv:1904.00083 [quant-ph]].
  %%CITATION = doi:10.3390/universe5040092;%%



  
  %\cite{Mukhanov:1981xt}
\bibitem{Mukhanov:1981xt} 
  V.~F.~Mukhanov and G.~V.~Chibisov,
  %``Quantum Fluctuations and a Nonsingular Universe,''
  JETP Lett.\  {\bf 33}, 532 (1981)
  [Pisma Zh.\ Eksp.\ Teor.\ Fiz.\  {\bf 33}, 549 (1981)].
  %%CITATION = JTPLA,33,532;%%



  
%\cite{Guth:1982ec}
\bibitem{Guth:1982ec} 
  A.~H.~Guth and S.~Y.~Pi,
  %``Fluctuations in the New Inflationary Universe,''
  Phys.\ Rev.\ Lett.\  {\bf 49}, 1110 (1982).
  %doi:10.1103/PhysRevLett.49.1110
  %%CITATION = doi:10.1103/PhysRevLett.49.1110;%%



  
%\cite{Hawking:1982cz}
\bibitem{Hawking:1982cz} 
  S.~W.~Hawking,
  %``The Development of Irregularities in a Single Bubble Inflationary Universe,''
  Phys.\ Lett.\  {\bf 115B}, 295 (1982).
 % doi:10.1016/0370-2693(82)90373-2
  %%CITATION = doi:10.1016/0370-2693(82)90373-2;%%



  
%\cite{Starobinsky:1982ee}
\bibitem{Starobinsky:1982ee} 
  A.~A.~Starobinsky,
  %``Dynamics of Phase Transition in the New Inflationary Universe Scenario and Generation of Perturbations,''
  Phys.\ Lett.\  {\bf 117B}, 175 (1982).
  %doi:10.1016/0370-2693(82)90541-X
  %%CITATION = doi:10.1016/0370-2693(82)90541-X;%%




  
%\cite{Bardeen:1983qw}
\bibitem{Bardeen:1983qw} 
  J.~M.~Bardeen, P.~J.~Steinhardt and M.~S.~Turner,
  %``Spontaneous Creation of Almost Scale - Free Density Perturbations in an Inflationary Universe,''
  Phys.\ Rev.\ D {\bf 28}, 679 (1983).
 % doi:10.1103/PhysRevD.28.679
  %%CITATION = doi:10.1103/PhysRevD.28.679;%%





%\cite{Guth:1980zm}
\bibitem{Guth:1980zm} 
  A.~H.~Guth,
  %``The Inflationary Universe: A Possible Solution to the Horizon and Flatness Problems,''
  Phys.\ Rev.\ D {\bf 23}, 347 (1981).
%  [Adv.\ Ser.\ Astrophys.\ Cosmol.\  {\bf 3}, 139 (1987)].
 % doi:10.1103/PhysRevD.23.347
  %%CITATION = doi:10.1103/PhysRevD.23.347;%%



  
%\cite{Linde:1981mu}
\bibitem{Linde:1981mu} 
  A.~D.~Linde,
  %``A New Inflationary Universe Scenario: A Possible Solution of the Horizon, Flatness, Homogeneity, Isotropy and Primordial Monopole Problems,''
  Phys.\ Lett.\  {\bf 108B}, 389 (1982).
%  [Adv.\ Ser.\ Astrophys.\ Cosmol.\  {\bf 3}, 149 (1987)].
  %doi:10.1016/0370-2693(82)91219-9
  %%CITATION = doi:10.1016/0370-2693(82)91219-9;%%




%\cite{Albrecht:1982wi}
\bibitem{Albrecht:1982wi} 
  A.~Albrecht and P.~J.~Steinhardt,
  %``Cosmology for Grand Unified Theories with Radiatively Induced Symmetry Breaking,''
  Phys.\ Rev.\ Lett.\  {\bf 48}, 1220 (1982).
%  [Adv.\ Ser.\ Astrophys.\ Cosmol.\  {\bf 3}, 158 (1987)].
  %doi:10.1103/PhysRevLett.48.1220
  %%CITATION = doi:10.1103/PhysRevLett.48.1220;%%



  
 %\cite{Guth:1985ya}
\bibitem{Guth:1985ya} 
  A.~H.~Guth and S.~Y.~Pi,
  %``The Quantum Mechanics of the Scalar Field in the New Inflationary Universe,''
  Phys.\ Rev.\ D {\bf 32}, 1899 (1985).
 % doi:10.1103/PhysRevD.32.1899
  %%CITATION = doi:10.1103/PhysRevD.32.1899;%%





%\cite{Grishchuk:1989ss}
\bibitem{Grishchuk:1989ss} 
  L.~P.~Grishchuk and Y.~V.~Sidorov,
  %``On the Quantum State of Relic Gravitons,''
  Class.\ Quant.\ Grav.\  {\bf 6}, L161 (1989).
%  doi:10.1088/0264-9381/6/9/002
  %%CITATION = doi:10.1088/0264-9381/6/9/002;%%





%\cite{Grishchuk:1990bj}
\bibitem{Grishchuk:1990bj} 
  L.~P.~Grishchuk and Y.~V.~Sidorov,
  %``Squeezed quantum states of relic gravitons and primordial density fluctuations,''
  Phys.\ Rev.\ D {\bf 42}, 3413 (1990).
%  doi:10.1103/PhysRevD.42.3413
  %%CITATION = doi:10.1103/PhysRevD.42.3413;%%




%\cite{Albrecht:1992kf}
\bibitem{Albrecht:1992kf} 
  A.~Albrecht, P.~Ferreira, M.~Joyce and T.~Prokopec,
  %``Inflation and squeezed quantum states,''
  Phys.\ Rev.\ D {\bf 50}, 4807 (1994)
%  doi:10.1103/PhysRevD.50.4807
  [astro-ph/9303001].
  %%CITATION = doi:10.1103/PhysRevD.50.4807;%%



  
 %\cite{Polarski:1995jg}
\bibitem{Polarski:1995jg} 
  D.~Polarski and A.~A.~Starobinsky,
  %``Semiclassicality and decoherence of cosmological perturbations,''
  Class.\ Quant.\ Grav.\  {\bf 13}, 377 (1996)
%  doi:10.1088/0264-9381/13/3/006
  [gr-qc/9504030].
  %%CITATION = doi:10.1088/0264-9381/13/3/006;%%




%\cite{Zurek:2003zz}
\bibitem{Zurek:2003zz} 
  W.~H.~Zurek,
  %``Decoherence, einselection, and the quantum origins of the classical,''
  Rev.\ Mod.\ Phys.\  {\bf 75}, 715 (2003)
  %doi:10.1103/RevModPhys.75.715
  [quant-ph/0105127].
  %%CITATION = doi:10.1103/RevModPhys.75.715;%%



  
%\cite{Schlosshauer:2003zy}
\bibitem{Schlosshauer:2003zy} 
  M.~Schlosshauer,
  %``Decoherence, the Measurement Problem, and Interpretations of Quantum Mechanics,''
  Rev.\ Mod.\ Phys.\  {\bf 76}, 1267 (2004)
 % doi:10.1103/RevModPhys.76.1267
  [quant-ph/0312059].
  %%CITATION = doi:10.1103/RevModPhys.76.1267;%%



  
%\cite{Burgess:2006jn}
\bibitem{Burgess:2006jn}
  C.~P.~Burgess, R.~Holman and D.~Hoover,
  %``Decoherence of inflationary primordial fluctuations,''
  Phys.\ Rev.\ D {\bf 77} (2008) 063534
  %doi:10.1103/PhysRevD.77.063534
  [astro-ph/0601646].
  %%CITATION = doi:10.1103/PhysRevD.77.063534;%%

  
   
  
%\cite{Kiefer:2006je}
\bibitem{Kiefer:2006je} 
  C.~Kiefer, I.~Lohmar, D.~Polarski and A.~A.~Starobinsky,
  %``Pointer states for primordial fluctuations in inflationary cosmology,''
  Class.\ Quant.\ Grav.\  {\bf 24}, 1699 (2007)
%%  doi:10.1088/0264-9381/24/7/002
  [astro-ph/0610700].
%  %%CITATION = doi:10.1088/0264-9381/24/7/002;%%

  
  
  
%\cite{Burgess:2014eoa}
\bibitem{Burgess:2014eoa} 
  C.~P.~Burgess, R.~Holman, G.~Tasinato and M.~Williams,
  %``EFT Beyond the Horizon: Stochastic Inflation and How Primordial Quantum Fluctuations Go Classical,''
  JHEP {\bf 1503}, 090 (2015)
  %doi:10.1007/JHEP03(2015)090
  [arXiv:1408.5002 [hep-th]].
  %%CITATION = doi:10.1007/JHEP03(2015)090;%%



  
%\cite{Burgess:2015ajz}
\bibitem{Burgess:2015ajz} 
  C.~P.~Burgess, R.~Holman and G.~Tasinato,
  %``Open EFTs, IR effects & late-time resummations: systematic corrections in stochastic inflation,''
  JHEP {\bf 1601}, 153 (2016)
  %doi:10.1007/JHEP01(2016)153
  [arXiv:1512.00169 [gr-qc]].
  %%CITATION = doi:10.1007/JHEP01(2016)153;%%



  
%\cite{Nelson:2016kjm}
\bibitem{Nelson:2016kjm} 
  E.~Nelson,
  %``Quantum Decoherence During Inflation from Gravitational Nonlinearities,''
  JCAP {\bf 1603}, 022 (2016)
  %doi:10.1088/1475-7516/2016/03/022
  [arXiv:1601.03734 [gr-qc]].
  %%CITATION = doi:10.1088/1475-7516/2016/03/022;%%



  
%\cite{Shandera:2017qkg}
\bibitem{Shandera:2017qkg} 
  S.~Shandera, N.~Agarwal and A.~Kamal,
  %``Open quantum cosmological system,''
  Phys.\ Rev.\ D {\bf 98}, no. 8, 083535 (2018)
 % doi:10.1103/PhysRevD.98.083535
  [arXiv:1708.00493 [hep-th]].
  %%CITATION = doi:10.1103/PhysRevD.98.083535;%%



  
%\cite{Martin:2018zbe}
\bibitem{Martin:2018zbe} 
  J.~Martin and V.~Vennin,
  %``Observational constraints on quantum decoherence during inflation,''
  JCAP {\bf 1805}, 063 (2018)
  %doi:10.1088/1475-7516/2018/05/063
  [arXiv:1801.09949 [astro-ph.CO]].
  %%CITATION = doi:10.1088/1475-7516/2018/05/063;%%




%\cite{Gong:2019yyz}
\bibitem{Gong:2019yyz} 
  J.~O.~Gong and M.~S.~Seo,
  %``Quantum non-linear evolution of inflationary tensor perturbations,''
  JHEP {\bf 1905}, 021 (2019)
%  doi:10.1007/JHEP05(2019)021
  [arXiv:1903.12295 [hep-th]].
  %%CITATION = doi:10.1007/JHEP05(2019)021;%%
  
  
  
  
%\cite{Lindblad:1975ef}
\bibitem{Lindblad:1975ef} 
  G.~Lindblad,
  %``On the Generators of Quantum Dynamical Semigroups,''
  Commun.\ Math.\ Phys.\  {\bf 48}, 119 (1976).
 % doi:10.1007/BF01608499
  %%CITATION = doi:10.1007/BF01608499;%%



  
%\cite{Banks:1983by}
\bibitem{Banks:1983by} 
  T.~Banks, L.~Susskind and M.~E.~Peskin,
  %``Difficulties for the Evolution of Pure States Into Mixed States,''
  Nucl.\ Phys.\ B {\bf 244}, 125 (1984).
  %doi:10.1016/0550-3213(84)90184-6
  %%CITATION = doi:10.1016/0550-3213(84)90184-6;%%




%\cite{Wigner:1932eb}
\bibitem{Wigner:1932eb} 
  E.~P.~Wigner,
  %``On the quantum correction for thermodynamic equilibrium,''
  Phys.\ Rev.\  {\bf 40}, 749 (1932).
  %doi:10.1103/PhysRev.40.749
  %%CITATION = doi:10.1103/PhysRev.40.749;%%



  
%\cite{Habib:1990hz}
\bibitem{Habib:1990hz} 
  S.~Habib,
  %``The classical limit in quantum cosmology. 1 Quantum mechanics and the Wigner function,''
  Phys.\ Rev.\ D {\bf 42}, 2566 (1990).
  %doi:10.1103/PhysRevD.42.2566
  %%CITATION = doi:10.1103/PhysRevD.42.2566;%%




%\cite{Habib:1990hx}
\bibitem{Habib:1990hx} 
  S.~Habib and R.~Laflamme,
  %``Wigner function and decoherence in quantum cosmology,''
  Phys.\ Rev.\ D {\bf 42}, 4056 (1990).
  %doi:10.1103/PhysRevD.42.4056
  %%CITATION = doi:10.1103/PhysRevD.42.4056;%%




%\cite{Martin:2015qta}
\bibitem{Martin:2015qta} 
  J.~Martin and V.~Vennin,
  %``Quantum Discord of Cosmic Inflation: Can we Show that CMB Anisotropies are of Quantum-Mechanical Origin?,''
  Phys.\ Rev.\ D {\bf 93}, no. 2, 023505 (2016)
  %doi:10.1103/PhysRevD.93.023505
  [arXiv:1510.04038 [astro-ph.CO]].
  %%CITATION = doi:10.1103/PhysRevD.93.023505;%%




%\cite{Hudson:1974}
\bibitem{Hudson:1974} 
  R.~Hudson,
  %``When is the wigner quasi-probability density non-negative?,''
  Rept.\ Math.\ Phys.\ {\bf 6}, 249 (1974).
%doi:10.1016/0034-4877(74)90007-X  




%\cite{Revzen:2004}
\bibitem{Revzen:2004}
 M.~Revzen, P.~A.~Mello, A.~Mann and L.~M.~Johansen,
  %``OBell's Inequality Violation (BIQV) with Non-Negative Wigner Function,''
  Phys.\ Rev.\ A {\bf 71},  022103 (2004)
  %doi:10.1103/PhysRevA.71.022103
  [arXiv:quant-ph/0405100].


 

%\cite{Martin:2017zxs}
\bibitem{Martin:2017zxs} 
  J.~Martin and V.~Vennin,
  %``Obstructions to Bell CMB Experiments,''
  Phys.\ Rev.\ D {\bf 96}, no. 6, 063501 (2017)
  %doi:10.1103/PhysRevD.96.063501
  [arXiv:1706.05001 [astro-ph.CO]].
  %%CITATION = doi:10.1103/PhysRevD.96.063501;%%




%\cite{Cheung:2007st}
\bibitem{Cheung:2007st} 
  C.~Cheung, P.~Creminelli, A.~L.~Fitzpatrick, J.~Kaplan and L.~Senatore,
  %``The Effective Field Theory of Inflation,''
  JHEP {\bf 0803}, 014 (2008)
 % doi:10.1088/1126-6708/2008/03/014
  [arXiv:0709.0293 [hep-th]].
  %%CITATION = doi:10.1088/1126-6708/2008/03/014;%%



  
%\cite{Prokopec:2010be}
\bibitem{Prokopec:2010be} 
  T.~Prokopec and G.~Rigopoulos,
  %``Path Integral for Inflationary Perturbations,''
  Phys.\ Rev.\ D {\bf 82}, 023529 (2010)
  %doi:10.1103/PhysRevD.82.023529
  [arXiv:1004.0882 [gr-qc]].
  %%CITATION = doi:10.1103/PhysRevD.82.023529;%%



  
%\cite{Gong:2016qpq}
\bibitem{Gong:2016qpq} 
  J.~O.~Gong, M.~S.~Seo and G.~Shiu,
  %``Path integral for multi-field inflation,''
  JHEP {\bf 1607}, 099 (2016)
  %doi:10.1007/JHEP07(2016)099
  [arXiv:1603.03689 [hep-th]].
  %%CITATION = doi:10.1007/JHEP07(2016)099;%%




%\cite{Gong:2017wgx}
\bibitem{Gong:2017wgx} 
  J.~O.~Gong and M.~S.~Seo,
  %``Consistency relations in multi-field inflation,''
  JCAP {\bf 1802}, 008 (2018)
%  doi:10.1088/1475-7516/2018/02/008
  [arXiv:1707.08282 [hep-th]].
  %%CITATION = doi:10.1088/1475-7516/2018/02/008;%%
  



%\cite{HongYi:1989}
\bibitem{HongYi:1989} 
  F.~Hong-Yi and J.~VanderLinde,
  %``Simple approach to the wave functions of one- and two-mode squeezed states,''
  Phys.\ Rev.\ A {\bf 39}, 1552 (1989).
 % doi:10.1103/PhysRevA.39.1552
 



%\cite{HongYi:1987zz}
\bibitem{HongYi:1987zz} 
  F.~Hong-Yi, H.~R.~Zaidi and J.~R.~Klauder,
  %``New approach for calculating the normally ordered form of squeeze operators,''
  Phys.\ Rev.\ D {\bf 35}, 1831 (1987).
 % doi:10.1103/PhysRevD.35.1831
  %%CITATION = doi:10.1103/PhysRevD.35.1831;%%




%\cite{Bloch:1937pw}
\bibitem{Bloch:1937pw} 
  F.~Bloch and A.~Nordsieck,
  %``Note on the Radiation Field of the electron,''
  Phys.\ Rev.\  {\bf 52}, 54 (1937).
  %doi:10.1103/PhysRev.52.54
  %%CITATION = doi:10.1103/PhysRev.52.54;%%







\end{thebibliography}
\end{document}